\documentclass[11pt,a4paper]{article}
\pdfoutput=1
\usepackage{myjheppub}
\usepackage[T1]{fontenc}
\usepackage{latexsym,amsfonts,amsmath,amssymb}
\usepackage{amsthm}
\usepackage{mathtools}
\usepackage{bbm}
\usepackage{graphicx}
\usepackage{color}

\usepackage[normalem]{ulem}
\usepackage{comment}
\usepackage{slashed}
\usepackage{bm}
\usepackage{hhline}

\usepackage{tabu}

\usepackage{physics}

\usepackage{float}

\usepackage{tikz}

\usepackage{hyperref}

\newcommand {\be} {\begin{equation}}
\newcommand {\ee} {\end{equation}}
\newcommand {\bea} {\begin{eqnarray}}
\newcommand {\eea} {\end {eqnarray}}

\newcommand{\CN}{\mathcal{N}}
\newcommand{\CH}{\mathcal{H}}

\newcommand{\ndt}{\noindent}

\renewcommand{\=}{\; = \;}

\newcommand{\p}{\partial}

\newcommand{\wt}{\widetilde}

\newcommand{\fl}{c}
\newcommand{\rch}{r}

\title{Localization and wall-crossing of giant graviton expansions in AdS$_5$}

\author{Giorgos Eleftheriou${}^1$, Sameer Murthy${}^{1}$, Mart\'i Rossell\'o${}^{\,2}$}

\affiliation{${}^1$ Department of Mathematics, King's College London, The Strand, London WC2R 2LS, UK}
\affiliation{${}^2$ Institute of Mathematics, Academia Sinica, Taipei, Taiwan}

\emailAdd{geleftheriou4@gmail.com}
\emailAdd{marti@gate.sinica.edu.tw}
\emailAdd{sameer.murthy@kcl.ac.uk}

\abstract{
The $\frac12$-BPS indices of $\mathcal{N}=4$ Super Yang-Mills theory with unitary, orthogonal, and symplectic groups 
all admit $q$-expansions suggesting an interpretation in terms of D-branes in the dual bulk AdS$_5$ string theories.
We present a derivation of these expansions in the corresponding bulk duals by quantizing the moduli space of $\frac12$-BPS giant 
gravitons using supersymmetric localization, extending and clarifying our study 
in~\href{https://arxiv.org/abs/2312.14921}{arxiv:2312.14921}.
We perform a detailed analysis of the one-loop fluctuations around the maximal giants (the fixed points), and show how the Hamiltonian analysis 
is recovered from the functional integral for the equivariant index. 
We show that the analytic continuation for these giant graviton expansions observed in the literature maps precisely to a wall-crossing phenomenon for the index.
In the case of orthogonal and symplectic gauge groups, the~$\mathbb{Z}_2$ quotient in the bulk 
leads to a corresponding projection in the $q$-expansion. 
Additional terms in the expansion related to the Pfaffian operator 
arise from topologically stable branes in the bulk dual on AdS$_5 \times \mathbb{RP}^5$. 
}

\begin{document}

\maketitle

\section{Introduction and summary of results}

The giant graviton expansion (GGE) refers to a certain type of power series expansion of superconformal 
indices of superconformal gauge theories~\cite{Arai:2019xmp,Imamura:2021ytr,Gaiotto:2021xce,Murthy:2022ien}. 
Each term in the series is itself a power series and is interpreted as the contribution of 
multi-determinant-type operators and their perturbations in the corresponding gauge theory. 
Such a determinantal expansion is seen only when we work at finite rank~$N$. 
Further, the signs of the terms in the series typically alternate.
These two aspects suggest an inclusion-exclusion principle for the index: one starts with 
a calculation at infinite~$N$, then removes finite-$N$ relations, puts back any overcounted 
states due to relations between relations, etc.

Such expansions were first discovered numerically to small orders in explorations of gauge theory indices,
and multiple versions of GGEs have been proposed for the same indices. 
When the series arises as a unitary matrix model, one can prove the existence of such an expansion 
and provide an explicit formula for each term~\cite{Murthy:2022ien,Liu:2022olj,Eniceicu:2023uvd,Eniceicu:2023cxn, Ezroura:2024wmp,Chen:2024cvf}. 
In this case, the determinantal nature of this expansion as well as the inclusion/exclusion principle  
are both explained~\cite{Murthy:2022ien} in terms of an underlying auxiliary free-fermion system studied in~\cite{BorOk}. 
This expansion has also been related to eigenvalue tunnelling~\cite{Eniceicu:2023uvd,Eniceicu:2023cxn,Chen:2024cvf}.

\smallskip

Our interest in this paper is in the interpretation of these $q$-expansions from the 
point of view of string theory in the holographic dual AdS$_{d+1}$ string/M-theory. 
The determinant-type operator is interpreted as a wrapped brane (giant graviton) in the bulk. 
The series of papers~\cite{Arai:2018utu,Arai:2019xmp,Arai:2020qaj,Arai:2020uwd,Fujiwara:2021xgu,
Imamura:2021ytr,Imamura:2022aua,Fujiwara:2023azx} has proposed a very interesting 
quiver-type theory using particular types of bulk branes, which 
reproduces the known expansions after an analytic continuation. 
One-loop calculations around certain symmetric single brane solutions 
have been shown to agree with the corresponding proposed giant-graviton expansions \cite{Beccaria:2023cuo, 
Beccaria:2023sph, Lee:2023iil,
Beccaria:2024vfx, 
Gautason:2024nru, Hatsuda:2024uwt, Lee:2024hef}. 
In another direction, there is evidence~\cite{Deddo:2024liu} for a different bulk interpretation, 
in terms of fluctuations of gravitons around backreacted brane (bubbling) 
geometries~\cite{Lin:2004nb} for~$\frac12$-BPS index in~$\CN=4$ Super Yang-Mills (SYM) 
theory.\footnote{There are, thus, three resulting interpretations of the GGE: in the boundary theory,
in terms of branes in AdS space, and in terms of bubbling geometries in AdS space. 
The first two routes use open string theory and the last one closed string theory---reflecting 
the open-closed-open triality~\cite{Gopakumar:2022djw}.
}

\smallskip

A complete and general bulk understanding, however, remains to be achieved. 
The approach we follow in this paper is to directly quantize the 
allowed supersymmetric brane configurations in the bulk, and calculate 
the contribution of any given brane state to the supersymmetric index. 
Typically, supersymmetric brane solutions in AdS$\times S$ have 
continuous parameters leading to a moduli space with multiple branches~\cite{Mikhailov:2000ya}. 
The moduli space of branes preserving a large amount of supersymmetry was 
quantized in the semiclassical limit in~\cite{Beasley:2002xv,Biswas:2006tj}. 
Our point of view here is that the GGE can be obtained from the bulk by 
appropriately quantizing the moduli space at finite~$N$.

\smallskip

In a previous paper~\cite{Eleftheriou:2023jxr}, we performed such a quantization 
for the simplest situation, namely the $\frac12$-BPS index in~$\CN=4$ 
SYM theory with~$U(N)$ gauge group, from the 
functional integral perspective.
This index is defined as a trace on the $\frac12$-BPS Hilbert space with the 
insertion of the fermion number $(-1)^F$, and graded by one of the Cartan 
generators~$R$ of the~$su(4)$ R-symmetry algebra of the theory, 
\begin{equation}
\label{eq:INtrace}
    {I_{U(N)}}(q)  \= 
     {\rm Tr}_{\CH^N_{\frac12\text{-BPS}}} \, 
    (-1)^F \, q^{R} \,.
\end{equation}
Here~$R$ takes non-negative values, and $q = e^{2 \pi i \tau}$ is taken to 
obey~$\vert q \vert < 1$, i.e.~$\text{Im}(\tau)>0$, for convergence of the above series.
The index is given by the expression~$I_{U(N)}(q)= 1/(q)_N$ in terms of 
the~$q$-Pochhamer symbol~$(q)_n=\prod_{i=1}^n (1-q^i)$. 
In this case, there is a simple analytic formula of the GGE form that is easily derived 
(see e.g.~Appendix~A of~\cite{Eleftheriou:2023jxr}), 
\begin{equation}
\label{eq:GGE_UN}
    {I_{U(N)}}(q)  \= 
    \frac{1}{(q)_N} 
    \= \frac{1}{(q)_\infty}\sum_{m=0}^\infty (-1)^m \; q^{mN} \, \frac{1}{(q)_m} \, q^{\binom{m+1}{2}} \,.
\end{equation}
The GGE~\eqref{eq:GGE_UN} for the~$U(N)$ gauge group and related GGEs presented below for 
orthogonal and symplectic groups are the focus of this paper. 

In~\cite{Eleftheriou:2023jxr} we expressed the $\frac12$-BPS 
index as an integral over the moduli space of~$\frac12$-BPS giants in~AdS$_5 \times S^5$,
and calculated it using the technique of supersymmetric localization.
This procedure leads to a term-by-term explanation of the formula~\eqref{eq:GGE_UN}. 
In the present paper we develop the localization calculation of the integral over the bulk giants in more depth.
In particular, we present a detailed calculation of the one-loop determinant
around the fixed point of the supercharge, 
thus clarifying some of the points in~\cite{Eleftheriou:2023jxr}. 
Our analysis relates the analytic continuation of the GGE observed in the literature~\cite{Arai:2019xmp,Gaiotto:2021xce} 
to the wall-crossing phenomenon of supersymmetric indices. 
Further, we generalize our method and results to $\CN=4$ SYM theories with orthogonal and 
symplectic gauge groups. 
The difference in the expansions for the different gauge groups is 
precisely accounted for by the details of the 
branes in the bulk duals of these theories.
In the rest of this introductory section we expand on these points in some detail.

\medskip

\ndt {\bf The localized integral over moduli space} \\
The ~$\frac12$-BPS giants in~AdS$_5 \times S^5$ wrap an~$S^3$ and rotate around 
a circle of~$S^5$ at the speed of light, and the corresponding angular momentum is identified 
with the~R-charge.\footnote{The conserved angular momentum is a sum of the kinetic angular 
momentum and the field angular momentum, the latter arising from the Chern-Simons 
coupling of the D-brane to the RR-fields.} 
The moduli space consists of two branches in which the D3-branes wrap $S^3 \subset S^5$ 
(giants) and $S^3 \subset$~AdS$_5$ (dual giants), respectively. 
The integral localizes to the fixed-point locus of a supercharge~$Q$, which (for one brane) 
is a single point in the moduli space,  corresponding to the so-called maximal 
giant.\footnote{The maximal giant stays at the center of AdS$_5$ and wraps a maximal $S^3 \subset S^5$
and, correspondingly, the circle around which they rotate  shrinks to zero size. In this 
case, all the angular momentum comes from the Chern-Simons coupling of the brane to the background RR-field.}
One can have an arbitrary number~$m=0,1,2,\dots$ of maximal giants preserving supersymmetry.

\smallskip

The contribution of the fixed points to the path integral is given by the quadratic fluctuation 
determinant of a $Q$-exact deformation around~$m$ maximal giants.
The quadratic fluctuation determinant factorizes into three pieces: the fluctuations of 
supergravitons in~AdS$_5 \times S^5$, a flux line carrying the background R-charge 
of~$m$ giants, and the (open string) fluctuations of the wrapped branes. 
The first two factors are easily calculated to be~$1/(q)_\infty$
and~$q^{mN}$, respectively, accounting for those factors in~\eqref{eq:INtrace}.

The open string fluctuations of the branes are particularly interesting. 
The bosonic fluctuation analysis can be done entirely in the DBI+CS action of the D-branes~\cite{Grisaru:2000zn}. 
While the fluctuations of generic giants (as well as all dual giants) are gapped~\cite{Das:2000fu}, 
the fluctuations of the maximal giants have gapless modes that are governed by the 
Landau Hamiltonian describing a charged particle in a magnetic field~\cite{Eleftheriou:2023jxr}. 
The full spectrum of the supersymmetric action, 
has been derived recently in~\cite{Gautason:2024nru}
(see~\cite{Imamura:2021ytr} for an earlier derivation exploiting the symmetries preserved by the brane). 
Using this result, we show in Section~\ref{sec:BPSbranes} that the  calculation of the 
supersymmetric index of all the brane fluctuations indeed reduces to the index of a supersymmetric version of 
the Landau problem, justifying the use of this truncated theory in~\cite{Eleftheriou:2023jxr}.

\smallskip

We calculate the index of the super Landau problem in the Hamiltonian formalism as well as 
in the Euclidean functional integral formalism. 
The Hamiltonian index is captured by the supersymmetric ground states of this theory, 
i.e.~the lowest Landau level.
In the Euclidean functional integral approach the index is given by the quadratic fluctuation 
determinant of the~$Q$-exact deformation mentioned above. 
This leads to the standard relation between character-valued indices and equivariant cohomology, 
which exactly reproduces the answer found from the Hamiltonian formalism. 
The calculation for~$m>1$ giants is a matrix version of the single brane quantum mechanics, and 
it precisely accounts for the remaining factor of~$(-1)^m \, q^{m(m+1)/2}/(q)_m$ 
in~\eqref{eq:GGE_UN}.

The above calculation involves an important subtlety which requires further discussion.
The supersymmetric ground states, and therefore the index,  of the super Landau problem 
for a fixed charge jump when the magnetic field~$B$ changes sign.
Interestingly, the grand-canonical index is an 
analytic function~\cite{Witten:1983ux,Stone:1988fu}, although its Fourier coefficients jump 
across the wall at~$B=0$. 
As we discuss in Section~\ref{sec:BPSbranes}, 
on one side of the wall we have the physical fluctuations of the brane, 
while on the other side we have the fluctuations of the~$Q$-exact 
deformation corresponding to a convergent GGE. 
This provides a physical explanation of the sign of the R-charge on the supersymmetric 
ground states that was chosen without a first-principles justification in~\cite{Eleftheriou:2023jxr} 
in order to obtain the GGE.\footnote{We thank Y.~Imamura, J.~H.~Lee, and 
D.~Stanford for illuminating discussions regarding this issue.}  
The $Q$-exact deformation also effectively regulates the IR divergence occurring at the location 
of the wall. The $Q$-deformed path integral can be analytically continued across the wall in 
a well-defined manner to give the two expansions. 
This explains the analytic continuation of the GGE for the~$\frac12$-BPS index discussed 
in~\cite{Arai:2019xmp,Gaiotto:2021xce} and subsequent works.

\smallskip

Thus, remarkably, localization allows us to promote a probe calculation in the bulk 
to the exact finite-$N$ result.
In particular, it explains (a) why we should focus only on maximal giants, and 
(b) why the one-loop approximation suffices,  even at finite~$N$. 
In contrast, in a perturbation expansion, one needs to take into account the higher 
loops of brane fluctuations as well as the non-linear couplings of brane  fluctuations to the bulk.

\bigskip

\ndt {\bf \texorpdfstring{$SO/Sp$}{SO/Sp} gauge groups and orientifolds}

\smallskip

$\CN=4$ SYM theory with~$SO$ or~$Sp$ gauge groups can be constructed in string 
theory using a stack of D3-branes along with an orientifold O3-plane. 
Depending on the charge of the orientifold one obtains an orthogonal or symplectic group. 
For each resulting gauge group, the corresponding $\frac{1}{2}$-BPS index can be derived 
combinatorially by enumerating the generators of $\frac{1}{2}$-BPS gauge invariant operators.

For the $SO(2k+1)$ and~$Sp(k)$ groups, 
the list of generators of the $\frac{1}{2}$-BPS gauge invariant operators is obtained 
from that for $U(N)$ with~$k=\lfloor N/2 \rfloor$, by projecting onto generators with even charge.
The GGE for these groups is therefore given by the replacement~$q \mapsto q^2$ in 
the $U(N)$ formula, namely, 
\begin{equation}
\label{eq:gge-sooddintro}
    I_{Sp(k)}(q) \=  I_{SO(2k+1)}(q)\= \frac{1}{(q^2)_\infty}\sum_{m=0}^\infty \, 
    (-1)^m \; q^{2 mk} \;\frac{1}{(q^2)_m} \;  q^{2\binom{m+1}{2}} \,.
\end{equation}
The equality of the indices for the $\CN=4$ theories with \hbox{$SO(2k+1)$} and $Sp(k)$ gauge groups 
is to be expected since they are Montonen-Olive duals \cite{MONTONEN1977117,  Witten:1998xy}.

For $SO(2k)$, the GGE is more interesting: 
\begin{equation} 
\label{eq:gge-so2kintro}
   I_{SO(2k)}(q)  \= \frac{1}{(q^2)_\infty}\sum_{m=0}^\infty (-1)^m \; q^{2mk}\, 
   \frac{1}{(q^2)_m} \,q^{2\binom{m+1}{2}}\,\bigl(1+q^k \bigr) \,.
\end{equation}
The new term in the expansion~\eqref{eq:gge-so2kintro} compared to~\eqref{eq:gge-sooddintro}  
appears due to the Pfaffian operator
which is of degree $k$ in the fields. 
Note that for $SO(2k+1)$ the Pfaffian term is absent since the determinant vanishes. 
For $Sp(k)$ the Pfaffian term is also absent since the Pfaffian is not well-defined for a 
general element in the Lie algebra of $Sp(k)$. 
We discuss the GGE for the different gauge groups in the boundary theory in more detail 
in Section~\ref{sec:boundary-indices}.

\medskip

The basic idea of the corresponding bulk analysis is as follows. 
The holographic duals of the~$SO$ and~$Sp$ theories are given by orientifold projections, 
which act in the bulk as \hbox{$S^5 \to S^5/\mathbb{Z}_2$} through antipodal identification, thus leading 
to AdS$_5 \times \mathbb{RP}^5$ bulk spacetime~\cite{Kakushadze:1998tr, 
Fayyazuddin:1998fb, Aharony:1998rm, Witten:1998xy}.\footnote{Giant graviton expansions 
of $SO/Sp$ gauge groups have also been discussed 
in~\cite{Fujiwara:2023bdc} by following the map to flat space branes~\cite{Arai:2019xmp} 
through the orientifold projection, and in~\cite{Hatsuda:2024lcc} using bubbling 
orientifolded geometries \cite{Mukhi:2005cv,Fiol:2014fla}.
}
The orientifold action reverses the orientation of the string worldsheet as one goes around 
a non-contractible loop in the~$\mathbb{RP}^5$. 
The different gauge groups for the boundary theories are distinguished by the different 
values of discrete torsion in the bulk~\cite{Witten:1998xy}.

D-branes in AdS$_5 \times S^5$ descend to D-branes in the orientifolded theories upon the 
appropriate bulk projection. The supersymmetric configurations which survive the projection 
are pairs of D-branes wrapping an $S^3\subset S^5$ rotating around a circle in $S^5$ at 
opposite points, with the orientifold projection identifying the antipodal points of their worldvolumes.  
In the theories dual to~$SO(2k+1)$ and~$Sp(k)$ gauge theories, these are the only supersymmetric D-branes. 
On a single pair of branes, the projection is purely geometric and keeps only the modes with 
even angular momenta on the~$S^5$ corresponding to the R-charge.
We explain this in Section \ref{sec:BPSbranesOrient}. 
On multiple pairs of branes, it involves an action on the matrix elements describing the D-brane 
worldvolumes which keeps only invariant modes after diagonalization. We explain this in 
Section~\ref{sec:conincidentbranes}. 
Together, the orientifold projection is implemented precisely by the replacement~$q \mapsto q^2$ 
in the expansion in the parent theory, which  reproduces~\eqref{eq:gge-so2kintro}. 

In the bulk dual to the~$SO(2k)$ theories, 
there is a new type of supersymmetric brane~\cite{Witten:1998xy} 
which is dual to the Pfaffian operator.
This brane exists because of the non-trivial $\mathbb{Z}_2$ homology~$H_3(\mathbb{RP}^5,\mathbb{Z})$, 
and it is rigid or ``topologically stable''~\cite{Aharony:2002nd} i.e.~it does not have any  fluctuations. 
It can, however, exist in its ground state along with any of the other supersymmetric branes,
which precisely explains the extra last term in~\eqref{eq:gge-so2kintro}.

\medskip

The plan of the rest of the paper is as follows. In Section~\ref{sec:BPSbranes} we present the 
details of the bulk localization for~$U(N)$ theories. In Section~\ref{sec:boundary-indices} 
we discuss the boundary indices for symplectic and orthogonal gauge groups.
In Section~\ref{sec:BPSbranesOrient} we describe the $\frac{1}{2}$-BPS moduli space of giants 
in AdS$_5\times\mathbb{RP}^5$ and discuss the  contributions to the GGE for the orthogonal 
group from two coincident maximal giants in the covering space and the topologically stable brane. 
In Section~\ref{sec:conincidentbranes} we extend the analysis to multiple branes in the orientifold theories. 
In an appendix we discuss the supersymmetry analysis in the Euclidean theory. 

\medskip

\noindent \textbf{Notation:} We denote by $Sp(k)$ the compact symplectic group of $2k \times 2k$ matrices. 
When $k$ and $N$ are used together for $SO(2k),\, SO(2k+1),\, Sp(k)$ and $U(N)$, 
they are always related by $k=\lfloor N/2 \rfloor$.

\section{Localization of \texorpdfstring{$\frac12$}{1/2}-BPS branes and their fluctuations \label{sec:BPSbranes}}

In this section we consider the $\frac{1}{2}$-BPS index in the~$U(N)$ theory
\begin{equation}
\label{eq:INtrace1}
    {I_{U(N)}}(q)  \= 
     {\rm Tr}_{\CH^N_{\frac12\text{-BPS}}} \, 
    (-1)^F \, e^{-\gamma(H-R)} \, q^{R} \,,
\end{equation}
where~$H$ is the Hamiltonian (conformal dimension) and~$R$ is one of the Cartan generators of 
the~$su(4)$ R-symmetry algebra. The index of the boundary gauge theory
only receives contributions from $\frac{1}{2}$-BPS states that are in the cohomology of a certain 
supercharge~$Q$ obeying~$Q^2 = H-R$. 
The states therefore obey~$H=R$, so that the trace~\eqref{eq:INtrace1} is independent of the 
parameter~$\gamma$ and can be written as~\eqref{eq:INtrace}.
Here the R-charge is chosen such that the only letter of the~$\CN=4$ SYM theory that 
contributes to the index are arbitrary excitations of one of the complex scalar matrices, which has~$H=R=1$. 
Below, we review the calculation of the index~\eqref{eq:INtrace1} as a bulk functional integral 
using localization as given in~\cite{Eleftheriou:2023jxr} and, in the process, elaborate on and 
clarify some points in that paper. 
In Section~\ref{sec:BPSbranesOrient} we repeat the calculation for the orientifold theories. 

\medskip

\ndt {\bf The bulk localization}

The starting point of the bulk dual of the index~\eqref{eq:INtrace} is the following  
integral over the moduli space of an abitrary number of~$\frac{1}{2}$-BPS giant gravitons in 
AdS$_5\times S^5$~\cite{Eleftheriou:2023jxr},  
\be \label{eq:PIgiant}
I_N^\text{bulk}(q) \= \sum_{m=0}^\infty \, \int_{\mathcal{M}_m} d\mu_m 
\int d \phi_m \, \exp \bigl(S_\text{brane+sugra} (\phi_m;\mu_m) \bigr) \, q^R \,.
\ee
Here~$\mu_m$ are the moduli of~$m$ $\frac{1}{2}$-BPS giants spanning the 
moduli space~$\mathcal{M}_m$, and $S_\text{brane+sugra}$ is the Euclidean 
action of the D3-brane coupled to the background supergravity.
The fluctuations of  supergravity fields and the collective coordinates of the branes 
in the ambient space are collectively denoted by $\phi_m$.

\medskip

The expression~\eqref{eq:PIgiant} is calculated as a Euclidean functional integral 
where the time coordinate describes a circle of size~$\gamma$. 
All fields, including the fermions, are periodic around the time circle, so as 
to implement the trace with~$(-1)^F$. 
We then choose a bulk supercharge~$Q$ that obeys the algebra~\hbox{$Q^2 = H-R$}.
The~$\frac{1}{2}$-BPS bulk configurations that contribute to the index~\eqref{eq:INtrace1} 
are given by the fixed points of~$Q$.

Next, we consider the operator insertion~$q^R$, which has a subtlety. 
Consider the index of a supercharge~$Q$ in a supersymmetric quantum theory. 
The basic index is the Witten index~$\text{Tr}_{\mathcal{H}} (-1)^F$
written as a trace over the full Hilbert space~\cite{Witten:1982df}. 
We can refine this trace by an additional operator insertion of the form~$q^{R}$ 
maintaining supersymmetry if the operator commutes with~$Q$. 
In such a situation we can also implement an additional twist in the Euclidean theory by 
turning on a holonomy for a background gauge field coupling to~$R$. 
However, in the case of the index~\eqref{eq:INtrace1}, the supercharge~$Q$ carries charge 
under~$R$ (equal to its charge under~$H$). 
This is the reason~\eqref{eq:INtrace1} is defined as a trace over the Hilbert subspace 
of~$\frac{1}{2}$-BPS states. In the  calculation of the functional integral, this means that 
we cannot turn on the background~$R$-symmetry gauge field 
proportional to the abitrary parameter~$\tau$ from the beginning in 
AdS$_5\times S^5$.\footnote{In the twisted theory~\cite{Gautason:2024nru}, 
this is seen by the fact that the background gauge field corresponding to the rotation~$R$, 
which is an off-diagonal component of the metric, takes a fixed value, corresponding to~$H=R$, 
in order to be supersymmetric.} Instead, as we see below, we will need to first localize to 
the~$Q$-cohomology and then turn on the refinement corresponding to~$R$.

\medskip

The path integral is then calculated by supersymmetric localization. 
This is achieved by deforming the classical action by a term of the form~$QV$.
The integral localizes to the critical points of~$QV$.
The critical points are precisely the so-called maximal giants that are separately 
annihilated by~$H$ and~$R$. 
In particular, this kills the contribution of all the other giants as well as the dual giants that 
expand in the AdS space.  
The contribution of each fixed point to the integral reduces to the exponential of the classical 
action times the one-loop functional determinant of~$QV$. 

This determinant was calculated by Hamiltonian methods in~\cite{Eleftheriou:2023jxr}. 
Although the correct answer was obtained for the one-loop determinant in~\cite{Eleftheriou:2023jxr}, 
the sign of the R-charge used in localization is different from the charge of the physical brane 
fluctuations. In~\cite{Eleftheriou:2023jxr}, the difference in reality conditions between the fluctuations 
contributing to the index and the physical fluctuations was noted, 
but a deeper explanation of how the R-charge of the fluctuations contributing to the index is related to 
the background R-charge was left unclear.
Below, we clarify this issue and present a more detailed and complete calculation of this determinant 
by Hamiltonian as well as path-integral methods.

\medskip

We begin the presentation with a brief review of the relevant aspects of giant gravitons in 
AdS$_5\times S^5$ in Section~\ref{sec:reviewGG}.
Next, in Section~\ref{sec:reviewLandau}, we recall 
how the small fluctuations of the maximal giant are summarized by the Landau problem, 
namely a two-dimensional particle in a constant transverse magnetic field. 
Then, in Section~\ref{sec:susyLan}, we study a supersymmetric version of the Landau 
problem and evaluate its index in the Hamiltonian framework.
Finally, in Section~\ref{sec:funcint}, we discuss the evaluation of the functional 
integral~\eqref{eq:PIgiant} in the Euclidean functional integral framework,
and show how this ties together the previous calculations. 
In particular, we show how the determinant of the~$QV$ deformation in the 
D3-brane theory on the sphere reduces to the calculation of the index in the supersymmetric 
Landau problem.

\subsection{Review of giant gravitons in  \texorpdfstring{AdS$_5\times S^5$}{AdS5xS5}  \label{sec:reviewGG}}

We use the same parametrization for the initial background AdS$_5\times S^5$ 
as in~\cite{Eleftheriou:2023jxr},
\be
 ds^2  \=  ds^2_{\text{AdS}_5} +  ds^2_{S^5} \,, 
\ee
with the respective line elements given by 
\begin{equation}
    ds^2_{\text{AdS}_5} \= - \Bigl(1 + \frac{r^2}{L^2}  \Bigr) \, dt^2 + 
    \Bigl(1 + \frac{r^2}{L^2}  \Bigr)^{-1} \, dr^2 + r^2 \, d\Omega_3^2 \,,
\end{equation}
and 
\begin{equation} \label{eq:s5metric}
    ds^2_{S^5} \= L^2 \bigl( d\theta^2 + \cos^2\theta \, d\phi^2 +\sin^2\theta \, d\Omega_3^2 \, \bigr) \,,
\end{equation}
where~$\theta \in [0,\frac{\pi}{2}]$,~$\phi\in[0,2\pi]$ and~$L$ is the radius of AdS$_5$ as 
well as~$S^5$. We use the following angular coordinates for the~$S^3$ inside~$S^5$, 
\begin{equation} \label{eq:S3-metric}
    d\Omega_3^2 \= d\chi_1^2 + \sin^2\chi_1 \, \bigl(d\chi_2^2+\sin^2\chi_2 \, d\chi_3^2 \, \bigr) \,,
\end{equation}
where $\chi_1,\chi_2 \in[0,\pi]$ and $\chi_3\in[0,2\pi]$. The 5-form field strength on $S^5$ is 
proportional to the volume form~$ F_{5} = \frac{4}{L} \text{Vol}(S^5)$ and it carries~$N$ units 
of flux corresponding to $N$ D3-branes in flat space,
\begin{equation} \label{eq:5-form-flux-N}
    \frac{1}{2\kappa^2_{10}T_3}\int_{S^5} F_{5} \= N \,.
\end{equation}
Writing the gravitational coupling as~$2\kappa^2_{10} = (2\pi)^7g_s^2 \alpha'^4$ and the 
D3-brane tension $T_3$ as~$T_3^{-1} = (2\pi)^{3}\alpha'^2g_s$, the radius of $S^5$ in 
terms of the flux takes the form~$L^4 = 4\pi g_s \alpha'^2 N$.  
The bosonic action for a D3-brane in this background is given by
\begin{equation} \label{eq:action-d3}
    S_3 \=-T_3 \int d^4\sigma \, \sqrt{-g} \; + \; T_3\int  P[A^{(4)}] \,,
\end{equation}
where~$\sigma_i$, $i=0,1,2,3$ are the worldvolume coordinates,
$g$ is the pullback of the spacetime metric $G$, and $P[A^{(4)}]$ is the pullback of the 4-form, 
which on the~$S^5$ we write as
\begin{equation} \label{eq:4form}
    A_{\phi\chi_1\chi_2\chi_3} \= L^4 \sin^4\theta \sin^2 \chi_1\sin\chi_2 
    \= L^4 \sin^4\theta \sqrt{g_{S^3}} \,.
\end{equation}
Giant gravitons are spherical~D3-branes wrapping $S^3\subset S^5$ which rotate around a circle 
in the five-sphere. In the static gauge, $\sigma_i = \chi_i$, they are described by 
the following embedding,  
\begin{equation} \label{GGsol}
   r \= 0   \,, \qquad \dot\phi \= \frac{1}{L} \,, \qquad \theta \= \theta_0 \,.
\end{equation}
Such configurations are solutions to the equations of motion derived from the 
action~\eqref{eq:action-d3}, and preserve half of the background 
supersymmetries~\cite{Grisaru:2000zn, Hashimoto:2000zp}. 
The constant~$\theta_0 \in [0,\pi/2]$ parameterizes the moduli space of giants. 
The giant labelled by $\theta_0$ has radius equal to~$L\sin\theta_0$ and angular 
momentum $N\sin^2 \theta_0$. 
(The angular momentum of the brane is a combination of the mechanical angular momentum 
and the effect of coupling to the background RR field.) 
The value~$\theta_0=\pi/2$ defines the maximal giant, which has largest possible radius 
equal to the~$S^5$ radius and angular momentum equal to $N$.
At this value of~$\theta_0$ the transverse circle along which the giant rotates, parametrized 
by~$\phi$, shrinks to zero size. 

\medskip

\subsection{Review of small fluctuations of maximal giants and the Landau problem \label{sec:reviewLandau}}

We now consider the small fluctuations of the giants. In particular, we consider the quantum-mechanical
 theory obtained by reducing the theory on the D3-branes on the~$S^3$. 
The light modes on the brane are a gauge field, six scalars and fermions. 
Upon KK reduction on the sphere, we obtain an infinite number of quantum-mechanical modes, 
most of which have an energy gap of the scale of the sphere. 
In fact, at a generic point in moduli space of giants and dual giants all the modes are 
gapped~\cite{Das:2000fu}. However, the quadratic fluctuations of the maximal giants contains, 
in the bosonic sector, two scalar fields 
which are the fields transverse to the~$S^3$ in the tangent space of the~$S^5$, which do not 
have have a mass term in the Lagrangian~\cite{Eleftheriou:2023jxr}. 

For small fluctuations, these scalars can be parameterized in terms of the variables introduced 
above as 
\begin{equation}
\label{eq:rhovarphicoords}
    \rho \= \frac{\pi}{2} - \theta  \, , \qquad \dot{\varphi} \= \frac{1}{L} - \dot{\phi} \,.
\end{equation}
It is convenient to introduce the variables~$(t,\phi)= (T, \frac{1}{L} T - \varphi)$ where~$T$ 
is the time coordinate in the theory of fluctuations. The charges of the fluctuations on the brane 
are related to the charges in the~AdS$_5 \times S^5$ as
\be \label{eq:tTrel}
-i  \p_\varphi \=  + i \p_\phi \= - R\,, \qquad 
\wt H \= i L \, \p_T 
 \= i L\, \p_t - i \p_\varphi \= H - R \,.
\ee
We rename~$\frac{1}{L}T \to t$ for ease of notation, so that in the following 
presentation $i\partial_t$ generates~\hbox{$H-R$}. 

\smallskip

The Lagrangian governing the small fluctuations of the maximal giant takes the following 
simple form~\cite{Eleftheriou:2023jxr},  
\begin{equation} \label{eq:quadratic-Lagrangian-angular}
      \mathcal{L} \= 
    \frac{N}{L} \biggl( 
    \frac{1}{2}\, L^2  \rho^2  \dot{\varphi}^2 + \frac{1}{2}\, L^2  \dot{\rho}^2 + L \rho^2  \dot{\varphi} -L\,\dot{\varphi}
    \biggr)  \,.
\end{equation}
This is the Lagrangian of the Landau problem describing a particle in a two-dimensional plane 
in the presence of a constant transverse magnetic field, together with an infinitely thin solenoid 
centered at the origin due to the linear term proportional to $\dot{\varphi}$. 
In terms of the Cartesian coordinates 
\be \label{eq:Cartesiancoords}
x_1 + i x_2 \= L \rho \, e^{i \varphi} \,, \qquad 
x_1 - i x_2  \= L \rho \, e^{-i \varphi} \,,
\ee
one obtains
\begin{equation} \label{eq:Lag-landau-w-flux}
     \mathcal{L} \= \frac{N}{L} \biggl( 
    \frac{1}{2} \, \bigl(\dot{x}_1^2+  \dot{x}_2^2 \bigr) + \frac{1}{L}(x_1\dot x_2 - x_2\dot x_1) \Bigl(
    1-\frac{L^2}{x_1^2+x_2^2}    \Bigr)    \biggr) \,.
\end{equation}
The last term with $x_1^2+x_2^2$ in the denominator, corresponding to the solenoid flux term, 
accounts for a shift by~$N$ units in the angular momentum~\cite{Eleftheriou:2023jxr}.
We ignore it in the analysis below until the very end when 
we reintroduce it in the accounting of the charges.

After discarding the solenoid term and absorbing the overall factor $N/L$ in the definition of the Lagrangian,
we obtain the Landau Lagrangian in the symmetric gauge for a charged particle of unit 
charge,\footnote{In the Landau Lagrangian the overall scale~$N/L$ 
uniformly scales the magnetic field as well as the mass of the particle. 
Since the energies of the Landau levels only depend on the cyclotron frequency, which is given by 
the ratio of the magnetic field and the mass, 
the energies are independent of the overall scale. See~\cite{Eleftheriou:2023jxr} for the formulas 
including the mass and charge.}
\begin{equation} \label{eq:Lag-landau-cartesian}
   \mathcal{L}_\text{Lan} \=  \frac{1}{2} \, \bigl(\dot{x}_1^2+  \dot{x}_2^2 \bigr) 
  +  \dot{x}_i \, A_i   \,,
\end{equation}
where~$\dot{}$ denotes the time-derivative, 
and the gauge fields are given by   
\begin{equation} \label{eq:A12defs}
  A_1 \=-\frac{B}{2} x_2 \,, \qquad 
  A_2 \= \frac{B}{2} x_1 \,, \qquad 
   \frac{\partial A_2}{\partial x_1}-\frac{\partial A_1}{\partial x_2} \= B \= \frac{2}{L} \,.
\end{equation}
The conjugate variables obeying the canonical commutation relations are 
\be
p_i = \frac{\partial \mathcal{L}_\text{Lan}}{\partial \dot x_i} \,, \qquad 
  [x_i,p_j] \= i\delta_{ij} \,,
\ee
and the corresponding Hamiltonian is 
\be \label{eq:Hamiltonian-Landau-rescaled}
  H_\text{Lan}   \=  \frac{1}{2}\, \bigl(p_1^2+p_2^2\bigr) 
  + \frac{1}{2}\left(\frac{B}{2}\right)^2\bigl(x_1^2+x_2^2\bigr)-\frac{B}{2}\bigl(x_1 p_2-x_2 p_1\bigr) 
  \= \frac{1}{2}\bigl( \vec{p} - \vec{A} \bigr)^2 \,.
\ee

\bigskip

The spectrum of the Landau problem is well-known. 
First we rearrange the four phase-space coordinates $x_i, p_i$, $i=1,2$ into the 
following linear combinations, 
\begin{equation} 
     \pi_1 \= p_1+\frac{1}{2}Bx_2 \,, \qquad \pi_2 \= p_2-\frac{1}{2}Bx_1 \,,
\end{equation}
and 
\begin{equation}
    X_1 \= \frac12 x_1 + \frac{1}{B} p_2\,, \qquad X_2 \= \frac12 x_2 - \frac{1}{B}p_1 \,.
\end{equation}
The operators~$\pi_i$ correspond to the kinetic momentum and the operators~$X_i$ 
correspond to the position of the center of the classical orbits \cite{yoshioka2002quantum}.
The corresponding quantum operators satisfy the following commutation relations (with the 
other commutators vanishing),
\begin{equation} \label{eq:Piicomms}
    [\pi_1, \pi_2] \= i B\,, \qquad [X_1, X_2] \= -\frac{i}{B} \,.
\end{equation}
In terms of these operators the Hamiltonian is given by 
\begin{equation}
    H_{\mathrm{Lan}} \= \frac{1}{2}\left( \pi_1^2 +  \pi_2^2  \right) \,.
\end{equation}
From the first commutation relation in~\eqref{eq:Piicomms}, we see that it is simply the 
Hamiltonian of a harmonic oscillator. 
Introducing creation and annihilation operators we can express it as 
\begin{equation} \label{eq:Ham-creation-ann}
     H_\text{Lan}   \=   B \Bigl(a^\dagger a + \frac{1}{2}\Bigr) \,, \qquad \bigl[ \,a \, , \, a^\dagger \,\bigr]=1\,.   
\end{equation}

The energy levels of~$H_\text{Lan}$, called the Landau levels, are infinitely degenerate.
The degeneracy is broken by the occupation number of another harmonic oscillator, 
which we call the~$b$ oscillator. To see its meaning, we define 
\begin{equation} \label{eq:defR2}
    \mathcal{R}^2 \= X_1^2+X_2^2 \,, 
\end{equation}
which measures the distance from the center of the classical orbit to the origin. 
From the second commutation relation in~\eqref{eq:Piicomms}, we see that this 
is also the Hamiltonian of a harmonic oscillator, which we write in terms of creation 
and annihilation operators as 
\begin{equation}
    \mathcal{R}^2 \= \frac{2}{B} \Bigl( b^\dagger b + \frac{1}{2} \Bigr) \,, \qquad  \bigl[ \, b\, , \,b^\dagger \, \bigr] \= 1 \,.
\end{equation}
In the symmetric gauge, the canonical angular momentum takes the form 
\begin{equation} \label{eq:LR2rel}
    \widehat{L}  \=  x_1 \, p_2 - x_2 \, p_1 
    \= -\frac{1}{2B}(\pi_1^2 + \pi_2^2) + \frac{1}{2}B(X_1^2 +X_2^2) \=  -a^\dagger a +  b^\dagger b\,.
\end{equation}

\medskip

Thus the full spectrum is characterized by two quantum numbers as follows,
\begin{equation} \label{eq:landau-eigenvalues}
    H_\text{Lan} \ket{\, n \,,\,\ell \,}\ \= B\bigl(n +\frac{1}{2} \bigr)\ket{\, n \,,\,\ell \,}\,, \qquad 
    \widehat{L}\ket{\, n \,,\,\ell \,} \= (-n+\ell)\ket{\, n \,,\,\ell \,} \,,
\end{equation}
where $n,\ell = 0,1,\dots$. 
The ground states~$\ket{0,\ell}$, defining the Lowest Landau Level (LLL), play an 
important role in our analysis. 
The wavefunctions of the  ground states are given by
\begin{equation} \label{eq:wf-lll-zz}
    \psi_{0,\ell} \= C_\ell \, z^\ell \, e^{-|z|^2 B/4} 
    \= C_\ell \, \rho^\ell \, 
    e^{i\ell \varphi} \, e^{-\rho^2 B/4} \,,
\end{equation}
in the complex coordinates $z = x_1+ix_2$, $\bar{z} = x_1-ix_2$ and the radial coordinates~$z = \rho e^{i \varphi}$, 
with $C_\ell$ a normalization constant. 
In particular, after factoring out the Gaussian damping, the wavefunction is holomorphic. 
This fact is useful in the analysis of the spectrum of fluctuations of coincident giants in 
Section~\ref{sec:conincidentbranes}. 
The wavefunctions carry positive angular momentum (anti-clockwise)
in the~$\rho, \varphi$ plane.\footnote{Note that a classical (positively) charged particle in an 
upward-pointing B-field spins clockwise under the Lorentz force. 
This is explained in \cite{vanenk} as the fact that~$\ell$ does not correspond to the physical 
gauge-invariant rotation.} 

\bigskip

In the brane problem, the LLL correspond to the $\frac12$-BPS fluctuations of the maximal giant. 
Using the map~\eqref{eq:tTrel}, we see that 
they carry negative values of R-charge. This is consistent with the fact that the maximal giant 
has maximal R-charge in the sector of physical fluctuations. 
Note, however, that there is a~$\mathbb{Z}_2$ symmetry of orientation reversal in the plane, 
underlying the theory of fluctuations. 
In the absence of the magnetic field, states with opposite signs of angular momentum are degenerate. 
Turning on~$B$ breaks this symmetry, but the Landau problem by itself is well-defined for either 
sign of the~$B$-field. 
The Hamiltonian clearly depends only on~$|B|$, while the~$\widehat{L}$-values of the LLL 
are correlated with~$\text{sign}(B)$. 
It is a simple exercise to obtain the general expressions for either sign of~$B$, and we use 
some of these expressions in the following section. 

\bigskip 

\subsection{The supersymmetric Landau problem and the Hamiltonian index \label{sec:susyLan}}

In this subsection we consider the minimal supersymmetric extension of the quantum-mechanical 
theory~\eqref{eq:Lag-landau-cartesian} and 
calculate a certain refined supersymmetric index of this theory. 
As we discuss in the next subsection, this index is the result of the functional integral of 
a $Q$-exact deformation around the maximal giant. 
In this subsection we analyze the index in its own right and discuss the resulting wall-crossing 
phenomenon in the Hamiltonian formalism. 

The relevant fields are~$(x_1,x_2,\lambda_1,\lambda_2)$, where~$\lambda_1, \lambda_2$ are 
Majorana fermions. The supersymmetric theory is described by the following simple 
Lagrangian,\footnote{One can think of this Lagrangian as a dimensional reduction of a supersymmetric
Chern-Simons theory in three dimensions~\cite{Gaiotto:2007qi}. The bilinear coupling of the fermions to~$B$ 
is also familiar from this description. 
}
see e.g.~\cite{BenGeloun:2009zhb, Correa:2010wk, Ivanov:2007sf, Kim:2001tw}: 
\begin{equation} \label{eq:susyLan}
  \mathcal{L}_\text{SLan} \= \sum_{i=1}^2 \, \Bigl( \, \frac{1}{2} \dot{x}_i^2 
  -\frac{1}{2}i \dot{\lambda}_i \, \lambda_i  +  \dot{x}_i \, A_i  \Bigr) - i \, B \, \lambda_1 \lambda_2\,,
\end{equation}
with the gauge fields given by~\eqref{eq:A12defs}.
This Lagrangian is invariant under the following supersymmetry transformations~\cite{BenGeloun:2009zhb}, 
\begin{equation}
\label{eq:SUSY-transforms}
    \delta \, x_i \= \lambda_i \,,  \qquad \delta \, \lambda_i \= i  \, \dot{x}_i \,, \qquad  i=1,2\,,
\end{equation}
which obey the algebra 
\be \label{QQbarsusyfluc}
\delta^2 \= i \, \p_{t} \,.
\ee
Note that this supersymmetry algebra and the relation of the various generators in~\eqref{eq:tTrel} 
are consistent with the algebra~$Q^2 = H-R$ of the supercharge defining the index~\eqref{eq:INtrace1}.

\bigskip

The theory~\eqref{eq:susyLan} is a non-interacting sum of the bosonic 
theory~\eqref{eq:Lag-landau-cartesian} and the theory of massive fermions. 
The corresponding Hamiltonian is 
\begin{equation}
     H_{\mathrm{SLan}} \= H_{\mathrm{Lan}} 
     + \frac{i}{2}\,B \, [\lambda_1,\lambda_2] \,.
\end{equation}
The fermionic theory is quantized by imposing the canonical commutation relations arising 
from the Dirac bracket for the fermions $\{ \lambda_i, \lambda_j \} =  \delta_{ij}$.\footnote{The 
fermionic Lagrangian is linear in time derivatives and defines a constrained system and 
therefore one must use the Dirac bracket instead of the Poisson bracket for canonical quantization.} 
These anticommutations relations define a two-dimensional Clifford algebra, 
which can be represented in terms of the Pauli matrices as~$\lambda_i = \frac{1}{\sqrt{2}}\sigma_i $, 
$i=1,2$, acting on a two-state system. 
This gives $[\lambda_1,\lambda_2] = i\sigma_3$ and the Hamiltonian is
\begin{equation}
     H_{\mathrm{SLan}} \= H_{\mathrm{Lan}} 
     - \frac{1}{2}\,B \, \sigma_3 \,.
\end{equation}

\smallskip

Equivalently, we can introduce complex fermions which satisfy the following algebra of fermionic 
annihilation and creation operators, 
\begin{equation} \label{eq:fermionic-creation-an-comm-rel}
    \fl \= \frac{1}{\sqrt{2}}\bigl(\lambda_1 + i\lambda_2\bigr)\,, \qquad \fl^\dagger 
    \= \frac{1}{\sqrt{2}}\bigl(\lambda_1 - i\lambda_2\bigr)\,, \qquad \{ \fl, \fl^\dagger \} 
    \= \delta_{ij} \,.
\end{equation}
The fermionic 2-state Fock space arising from \eqref{eq:fermionic-creation-an-comm-rel} 
is written as $\ket{n_F}$, $n_F = 0,1$, where $\fl\ket{0} =0$ and $\fl^\dagger\ket{0}=\ket{1}$.
The full Hamiltonian and angular momentum operator $\widehat{L}$ take the following form, 
\begin{equation} \label{eq:HLabc}
     H_{\mathrm{SLan}} \= |B|\Bigl(a^\dagger a +\frac{1}{2}\Bigr) 
     +B\Bigl( \fl^\dagger \fl - \frac{1}{2} \Bigr)\,, \qquad \widehat{L} 
     \= \mathrm{sign}\,(B)\bigl(-a^\dagger a + b^\dagger b\bigr) - \fl^\dagger \, \fl \,.
\end{equation}
The spectrum of the Hamiltonian including the fermionic fields is 
\begin{equation} \label{eq:Hferspec}
        H_{\mathrm{SLan}} \ket{n,\ell,s} \= \Bigl( |B| \bigl(n+\tfrac{1}{2}\bigr) 
        + B \bigl(n_F - \tfrac{1}{2}\bigr)   \Bigr)\ket{n,\ell,n_F}\,, 
\end{equation}
where~$n, \ell = 0,1,\dots$ as before and 
the angular momentum operator has the following eigenvalues
\begin{equation} \label{eq:Lferspec}
        \widehat{L} \ket{n,\ell,s} \=  \Bigl( \text{sign}(B) \bigl(-n+\ell \bigr) -  n_F  \Bigr)\ket{n,\ell,n_F} \,.
\end{equation}
The fermion number operator~$F = \fl^\dagger \fl$ yields~$(-1)^F = (-1)^{n_F}$.

\medskip

In the above formulas, we have allowed for~$B$ to have either sign. 
While the bosonic Hamiltonian depends only on~$|B|$, the fermionic spin tends to align 
with the direction of~$B$. 
Note that the Hamiltonian is supersymmetric for either choice of~sign$(B)$. 
The supercharge takes the following form,\footnote{The 
different expressions depending on the sign of $B$ stem from the fact that to satisfy the 
commutation relations $[a,a^\dagger]=1$ the operators $a,a^\dagger$ must depend on 
the sign of $B$ as $a = \frac{1}{\sqrt{2|B|}}(\pi_1 +i\, \mathrm{sign}(B)\pi_2)$. }
\begin{equation} \label{eq:supercharge-Q}
    Q \= \frac{1}{\sqrt{2}} \left( \sigma_1\pi_1 + \sigma_2 \pi_2  \right) \= \begin{cases}
        \sqrt{|B|}\begin{pmatrix}
        0 & a^\dagger \\
        a & 0
    \end{pmatrix} \= \sqrt{|B|}\bigl( a^\dagger \fl + a\, \fl^\dagger\bigr)\; & \mathrm{for }\; B>0\,, \\
         \sqrt{|B|}\begin{pmatrix}
        0 & a \\
        a^\dagger & 0
    \end{pmatrix} \= 
    \sqrt{|B|}\bigl( a^\dagger \fl^\dagger + a\, \fl )\; & \mathrm{for }\; B<0\,.
    \end{cases}
\end{equation}
It satisfies $Q^\dagger = Q$ and 
\begin{equation}
    Q^2 \= H_{\mathrm{SLan}} \,,\qquad \, \bigl[ H_{\mathrm{SLan}} \,,\, Q \bigr] \= 0\, .
\end{equation}

\medskip 

The Witten index~$\mathrm{Tr}_{\mathcal{H}_\text{SLan}}(-1)^F e^{-\gamma H_{\mathrm{SLan}}}$ 
is not well defined due to the infinite degeneracy in every Landau level including the~LLL. 
Since $\widehat{L}$ commutes with the supercharges and the Hamiltonian,
\begin{equation}
    \bigl[ H_{\mathrm{SLan}} \,,\, \widehat{L}\, \bigr] \= 0\,,\qquad \bigl[Q \,, \,\widehat{L} \, \bigr] \= 0\,,
\end{equation}
one can refine the Witten index 
by the angular momentum to obtain a well-defined quantity. 
Since all states with $H_{\mathrm{SLan}}\neq 0$ are paired with opposite values of~$(-1)^F$, 
the trace over the Hilbert space gets reduced to a trace over the states of the LLL,
which are labelled by their~$\widehat{L}$-eigenvalue.

\bigskip

In order to relate this discussion to the trace~\eqref{eq:INtrace1}, we need to label the states by 
their $R$-eigenvalue. To this end, we re-introduce the solenoid flux term 
in~\eqref{eq:quadratic-Lagrangian-angular} which we had ignored until now. 
Its effect is to implement the shift~$\widehat{L} \to \widehat{L}-N$ in the angular 
momentum~\cite{Eleftheriou:2023jxr}, so that~$R = - i\partial_{\phi} = N-\widehat{L}$.
With the above considerations, we obtain the following expressions for the index of the 
fluctuations of the maximal giant, 
\be \label{eq:HSLanInd-R-trace}
\text{Tr}_{\mathcal{H}_\text{SLan}} \, (-1)^F \, e^{-\gamma H_\text{SLan}} \, 
q^{R} \=
\begin{cases}
q^N\sum\limits_{\ell = 0}^{\infty} \, q^{-\ell} \= \dfrac{q^N}{1-q^{-1}} \,, & \qquad B>0 \,, \qquad |q|>1 \,, \\\
\vspace{0.1cm} \\
q^N\sum\limits_{\ell = 0}^{\infty} \, q^{\ell+1} \= -\dfrac{q^{N+1}}{1-q} \,, & \qquad B<0 \,, \qquad |q|<1 \,. \
\end{cases}
\ee 

In the first case, the series is  convergent when~$|q|>1$, and is identified with the physical  
fluctuations of the brane which reduce the value of~$R$ compared to the ground state of the maximal giant. 
The overall constant in~$\widehat{L}$ is fixed by demanding that the ground state of the 
maximal giant has~$R=N$, so that 
the R-charges of the physical fluctuations take the integer values~$N,N-1,N-2,\dots$.\footnote{This 
same result is recovered in a different manner in~\cite{Arai:2019xmp}.  
When one expands around the maximal giant solution in a frame which is co-rotating with the giant, 
one recovers the Landau problem as above.
In contrast, an expansion around the stationary frame is considered in~\cite{Arai:2019xmp} and 
subsequent works. 
The supersymmetry algebra on these stationary branes is then cleverly mapped to the usual 
supersymmetry algebra of branes in flat space. 
Following this map leads to an expansion in negative values of the R-charge.}
The states contributing to the trace have~$n_F = 0$, which is consistent with the 
ground state and the physical fluctuations of the giant being bosonic.

In the second case, the series is convergent when~$|q|<1$, and corresponds to the 
regime of the GGE, where $R = N+1, N+2,\dots$. 
In this case, the fermion occupation number is~$n_F=1$, which produces the overall sign 
as well as the additional shift of the angular momentum by one unit.   
It is this latter regime with~$B<0$ that we choose in  the $Q$-deformation of the action in the 
localization calculation in order to reproduce the GGE, 
as we discuss in more detail in the next subsection.\footnote{The above discussion clarifies the 
relation between the R-charge and the angular momentum used in~\cite{Eleftheriou:2023jxr}. 
In particular, although the geometric R-charge~$-i \partial_\phi$ takes negative values on 
the physical fluctuations of the brane, it takes positive values on the fluctuation spectrum 
when~$B<0$.}

\medskip 

Note that, although the spectrum of the index~\eqref{eq:HSLanInd-R-trace} changes 
under a change of sign of~$B$,
the index regarded as an analytic function of~$q$ does not change. 
Indeed, the same model as the one we study above was studied in~\cite{Witten:1983ux, Stone:1988fu} 
where similar comments are made
about the analytic behavior.\footnote{Those papers discuss the ensemble with fixed eigenvalue 
under~$\widehat{L}$. This leads to the appearance of two bosonic harmonic oscillator quantum numbers
labelling the energy eigenstates, in contrast to our treatment, where the bosonic Hamiltonian is that 
of a  single harmonic oscillator.} 
This behaviour can be understood as an instance of the wall-crossing phenomenon, 
where the index changes discontinuously as $B$ changes sign. 
The wavefunctions of the ground states  must be annihilated by the supercharge $Q$. 
Using the position space representation of $\pi_1$ and $\pi_2$ 
in~\eqref{eq:supercharge-Q}, we see that 
\begin{equation} \label{eq:wavefunc-annihilated-q}
Q \, \psi(z,\bar{z}) \= 0 \quad \implies \quad     \psi(z,\bar{z}) \= \begin{pmatrix}
        \psi_1(z,\bar{z}) \\
        \psi_2(z,\bar{z})
    \end{pmatrix} \= \begin{pmatrix}
        f(z) \, e^{-\frac{B}{4}z\bar{z}} \\
         g(\bar{z}) \, e^{\frac{B}{4}z\bar{z}}
    \end{pmatrix} \,,
\end{equation}
where~$f$ and~$g$ are arbitrary holomorphic and antiholomorphic functions, respectively. 
Depending on sign$(B)$, only one of $\psi_1(z,\bar{z})$ and $\psi_2(z,\bar{z})$ is normalizable. 
This gives a physical model for the intriguing  observations~\cite{Imamura:2021ytr, 
Gaiotto:2021xce, Imamura:2022aua}  about the 
analytic continuation of the~$\frac12$-BPS index of giant gravitons. 

\medskip

To summarize the story so far, the index of the minimally supersymmetric Landau problem 
depends on the sign of the~$B$-field. 
The states with negative values of angular momentum contribute to the convergent GGE.
They are the BPS spectrum of the super-Landau Hamiltonian with~$B<0$, which should be 
identified as the~$Q$-deformation to the physical brane Lagrangian. 
In terms of the~$a$, $b$, and~$c$ oscillators, the BPS states are the eigenstates of 
the~$b$ oscillator in the fermionic state~$c^\dagger c =1$. 
The excitations of the bosonic~$a$ and fermionic~$c$ oscillators cancel out in the index.

The refinement of the index by~$q^{R}$ can be also be calculated by the Euclidean 
functional integral.
To do this we need to deform the supercharge so as to capture the effect of angular 
momentum~\cite{Witten:1983ux, Stone:1988fu} in the deformed action, so as to
obtain the equivariant cohomology. 
In the following subsection we discuss this deformation and show how precisely the  
index~\eqref{eq:HSLanInd-R-trace} of the super-Landau problem is recovered as the determinant of 
a~$Q$-exact deformation to the brane action.

\subsection{The Euclidean functional integral analysis \label{sec:funcint}}

In the Euclidean theory, we rotate the time coordinate to an imaginary direction~$t=-i t_E$ 
and make the identification~$t_E \sim t_E + \gamma$.
In order to preserve supersymmetry in this background, we need to implement a twist in 
the~$\phi$ direction by turning on an imaginary off-diagonal component of the metric.
This twisted background is exactly the  Euclidean continuation of the co-rotating frame for 
giants which rotate around the~$\phi$ circle, which shows that such 
branes preserve all the 16 supercharges. 
One can check this explicitly following the $\kappa$-symmetry arguments 
in \cite{Grisaru:2000zn}, as we discuss in Appendix \ref{sec:Appendix-killingspinors}.

Now we turn to the fluctuations of the maximal giants in the Euclidean theory. 
The spectrum of fluctuations of light fields on the D3-brane consists of three complex scalar fields, 
one vector field, and fermions on~$S^3 \times S^1_\gamma$~\cite{Gautason:2024nru}.\footnote{This 
coincides with the spectrum derived in~\cite{Imamura:2021ytr} 
using representations of the supersymmetry algebra.}  
We view this spectrum as an infinite number of quantum-mechanical variables depending on time. 
Since the problem is spherically symmetric, the quantum variables are labelled by 
only one quantum number~$k$, the level of the spherical harmonic on the~$S^3$, and have
a degeneracy~$d_k$. 
The corresponding eigenvalue of the Laplacian on~$S^3$ is called~$\omega_k$.  
Each field also carries an R-charge, which we denote by~$\rch$, corresponding to the rotation 
in the~$\phi$ direction. 
The frequencies of the quantum variables are  given by~$\omega_{k,\rch} = \omega_k + \rch$.
We can further expand each field in 
momentum-modes~$n_\gamma = 2 \pi n/\gamma$, $n \in \mathbb{Z}$ around the~$S^1_\gamma$. 
The eigenvalue of each such mode under the quadratic operator~$\mathcal{K}$ 
appearing in the action is given by~$\mathcal{K}_{k,r}=\omega_{k,r}^2 + n_\gamma^2$. 
In Table~\ref{table:spectrum} we summarize the 
spectrum of quadratic fluctuations of the fields living on the maximal giant.
\begin{table}[ht]
\renewcommand{\arraystretch}{1.1}
    \centering
    \begin{tabular}{ c|c|c|c|c } 
Field & Degeneracy 
$d_k$ &R-charge $\rch$ & 
$\mathcal{K}_{k,r}$ \\
\hline
Real scalars & $4 \times (k+1)^2 $ & 0  & $(k+1)^2 + n_\gamma^2$  \\
 & $1 \times (k+1)^2$ & $+1$ & $(k+2)^2 + n_\gamma^2$ \\
 & $1 \times (k+1)^2$ & $-1$ & $k^2 + n_\gamma^2$\\
 \hline
Fermions & $2 \times 2(k+1)(k+2)$  & $ +\frac{1}{2}$ & $\left(k+2\right)^2 + n_\gamma^2$\\ 
 & $2 \times 2(k+1)(k+2)$  & $- \frac{1}{2}$ & $\left(k+1\right)^2 + n_\gamma^2$\\ 
 \hline
Vectors & $1 \times 2(k+1)(k+3)$  & 0  & $(k+2)^2+n_\gamma^2$\\ 
\end{tabular}
    \caption{Spectrum of quadratic fluctuations of the Euclidean maximal giant~\cite{Gautason:2024nru}. 
    Here $k$ labels the level of the spherical harmonics on~$S^3$ of the various fields,   
    $d_k$ is the degeneracy of the field at each level $k=0,1,2,\dots$, $\rch$ its R-charge, and 
    $\mathcal{K}_{k,r}=\omega_{k,r}^2 + n_\gamma^2$, $n_\gamma = \frac{2 \pi n}{\gamma}$, 
    $n\in \mathbb{Z}$, is the eigenvalue of the field under the quadratic one-loop operator.}
    \label{table:spectrum}
\end{table}

\medskip

Let us consider the ratio of fermionic and bosonic determinants 
that is the result of  integrating out all the quadratic fluctuations of the theory. 
The logarithm of this quantity 
is given by 
\begin{equation} \label{eq:trace-log-one-loop}
    \frac{1}{2}\mathrm{Tr} \log \mathcal{K} 
    \= \frac{1}{2} \sum_{\Phi} (-1)^{F_{\Phi}}\sum_{n\in\mathbb{Z}}
    \sum_{k=0}^\infty d_k \log \bigl(\omega_{k,\rch}^2 + n_\gamma^2 \bigr)\,,
\end{equation}
where the sum over $\Phi$ represents the sum over the fields in Table \ref{table:spectrum}, 
$(-1)^{F_{\Phi}}$ is~$-1$ for fermions and~$+1$ 
for bosons, and~$\rch$ appearing in~$\omega_{k,\rch}$ is the charge of the field~$\Phi$. 
When we sum over the modes listed in Table \ref{table:spectrum}, we observe many cancellations 
in the sum~\eqref{eq:trace-log-one-loop}.  
In fact, all the modes in Table~\ref{table:spectrum} are paired by the supercharge 
except for one quantum-mechanical bosonic field with~$k=0$, $\rch=-1$ which is unpaired. 
The sum~\eqref{eq:trace-log-one-loop} reduces to the formal sum 
\begin{equation} \label{eq:trace-log-one-loop-after-cancellations}
    \frac{1}{2}\mathrm{Tr} \log \mathcal{K} \= \frac{1}{2} \sum_{n\in\mathbb{Z}} \log n_\gamma^2 \,,
\end{equation}
which is the contribution of the unpaired bosonic field. 
The~$n=0$ mode is a zero mode of the whole action and has to be treated separately.  

\smallskip

We can now identify the supersymmetric Landau system studied in Section~\ref{sec:susyLan} 
as a sector of the full brane theory. 
Explicitly, we keep the two bosonic fields labelled by~$k=0$, $\rch = \pm 1$ and two fermionic fields 
labelled by $k=0$, $\rch=\frac12$.
All other modes of the brane appear as boson-fermion pairs. 
In other words, the supersymmetric index of the brane fluctuations reduces to the supersymmetric 
index of the super-Landau problem. 
As consistent with the brane analysis, there is also a further cancellation in the super-Landau 
problem between the excitations of one of the bosonic modes and the two fermionic modes, so as 
to leave one unpaired quantum-mechanical bosonic field. 
However, we do not implement this cancellation and 
keep these fields in the theory at this stage.\footnote{In~\cite{Eleftheriou:2023jxr} we did not derive 
the fermionic spectrum of small fluctuations of the D-brane from first principles, and 
the original version of the paper contains an incorrect guess that there are four Majorana fermions in the surviving theory.
The above analysis fills this gap and corrects this error.}

The reason it is useful to study the full super-Landau problem---with the complex scalar, complex fermion, 
and the local coupling to the $B$-field---is that it captures all the symmetries of the brane that are relevant 
to the index.  
As we see below, this allows us to easily refine the index with the angular-momentum (R-charge) so 
that all the modes in the sum~\eqref{eq:trace-log-one-loop-after-cancellations} obtain an effective non-zero mass. 
Before studying the refined index, we check the above conclusion of the reduction of the index  
directly by studying the functional integral of the super-Landau problem. 
We also use this to set up our conventions for the path integral corresponding to the refined trace that we 
calculate below. 

\medskip

Upon Wick-rotation, 
the Lagrangian~\eqref{eq:susyLan} gives
\begin{equation} \label{eq:SLanEucAct}
S^E_\text{SLan} \= \int_0^\gamma dt_E \,  \mathcal{L}^E_\text{SLan} \,, 
\end{equation}
with
\begin{equation} \label{eq:SLanEuc}
   \mathcal{L}^E_\text{SLan} \= \sum_{i=1}^2 \, \Bigl( \, -\frac{1}{2} \dot{x}_i^2 
  +\frac{1}{2} \dot{\lambda}_i \, \lambda_i \Bigr) 
  + i \frac{B}{2} \bigl( -\dot{x}_1 x_2 + \dot{x}_2 \, x_1 \bigr) - i \, B \, \lambda_1 \lambda_2\,.
\end{equation}
The supersymmetry variations are as follows, 
\begin{equation}
\label{eq:EucSusyVars}
    \delta \, x_i \= \lambda_i \,,  \quad \delta \, \lambda_i \= -  \, \dot{x}_i 
     \qquad \Longrightarrow \qquad 
    \delta^2 \= - \, \p_{t_E} \,.
\end{equation}

Using the translational invariance around the circle, we can expand the real fields~$x_i(t_E)$, $\lambda_i(t_E)$ 
in Fourier modes as 
\be \label{eq:xlamFour}
x_i(t_E) \= \sqrt{\gamma} \sum_{n \in \mathbb{Z}} \, x^i_n \, e^{2 \pi i \, n \, t_E/\gamma} \,, \qquad 
\lambda_i(t_E) \= \sum_{n \in \mathbb{Z}} \, \lambda^i_n \, e^{2 \pi i \,n \, t_E/\gamma} \,.
\ee
The action of the theory in terms of the Fourier modes is~$S^E_\text{SLan} = S^E_\text{b}+S^E_\text{f}$ with
\be
\begin{split}
S^E_\text{b} & \= - \sum_{n\ge 1} \, 2\pi n \, \bigl( \, x^1_n \; x^2_n \,\bigr) 
\biggl(\, \begin{matrix} 2\pi n & - B \gamma \\ B\gamma & 2\pi n \end{matrix}\,  \biggr) 
\biggl(\begin{matrix}x^1_{-n} \\  x^2_{-n} \end{matrix} \biggr) \,, \\
S^E_\text{f} & \= i \sum_{n\ge 1} \, \bigl(\, \lambda^1_n \; \lambda^2_n \, \bigr) 
\biggl(\, \begin{matrix} 2\pi n & - B \gamma \\  B\gamma & 2\pi n \end{matrix} \, \biggr) 
\biggl(\, \begin{matrix}\lambda^1_{-n} \\  \lambda^2_{-n} \end{matrix} \, \biggr) 
- i B \gamma \, \lambda^1_0 \, \lambda^2_0 \,.
\end{split}
\ee

The functional integral governing the fluctuations of the super-Landau problem is 
\be \label{eq:SlanPI}
Z \= \int [Dx \, D\lambda] \, \exp \bigl( S^E_\text{SLan}(x_i(t_E), \lambda_i(t_E)) \bigr) \,,
\ee
with the measure given by 
\be
[Dx \, D\lambda] 
\= \mathcal{N} \prod_{i=1,2} \, dx^i_0 \, d\lambda^i_0 \, \prod_{n \ge 1} dx^i_n \, dx^i_{-n} \,   
d\lambda^i_n \, d \lambda^i_{-n} \,,
\ee
where the overall normalization constant is fixed at the end by physical considerations.

\medskip

Let us first consider the bosonic integrals.
The integrals over the non-zero modes~$x^i_n, x^i_{-n}$ yields 
the product of the corresponding free Gaussian integrals and  
\be 
\label{eq:bosdet}
\prod_{n=1}^\infty \, (2 \pi n)^{-2} \, \text{det}  
\biggl(\, \begin{matrix} 2\pi n & - B \gamma \\ B\gamma & 2\pi n \end{matrix}\,  \biggr)^{-1} \= 
\prod_{n=1}^\infty \, (2 \pi n)^{-2} \, \bigl( 4 \pi^2 n^2 + B^2\gamma^2 \bigr)^{-1} \,.
\ee  
The integral over the zero modes gives a factor of  
the (infinite) volume~$V_2$ of the~$(x_1,x_2)$-plane.

The fermionic integral is calculated similarly. 
The integrals over the non-zero modes~$\lambda^i_n$ result in 
\be \label{eq:ferdet}
\prod_{n=1}^\infty \, \text{det}  
\begin{pmatrix} 
2 \pi n & -  B \gamma \\
 B \gamma & 2 \pi n  
\end{pmatrix} \= 
\prod_{n=1}^\infty \, \bigl( 4 \pi^2 n^2 + B^2 \gamma^2 \bigr) \,.
\ee 
The numerical factors in the overall normalization of fermionic integrals are chosen 
to be consistent with the supersymmetric cancellation. 
The integral over the non-zero modes of the fermions in~\eqref{eq:ferdet} 
cancel the integral over the non-zero modes of the bosons in~\eqref{eq:bosdet}.  
The eigenvalues of these modes for~$B=2$ (with $L=1$) are precisely that of the boson 
with~$R=+1$ in Table~\ref{table:spectrum}. 
Finally, the integral over the zero-modes~$\lambda^i_0$ results in a factor 
of~$B \gamma$ in the numerator.

Putting together the various pieces we obtain
\be \label{eq:ZV2B}
Z \= \mathcal{N} \, V_2 \, B \gamma \, 
\prod_{n=1}^\infty \, \frac{1}{4 \pi^2 n^2} \,.
\ee 
The above formula explicitly shows that the unpaired modes contributing to the index 
correspond to those of one free boson, in accord with the Hamiltonian formalism in the 
previous subsection as well as the brane analysis~\eqref{eq:trace-log-one-loop-after-cancellations}. 
(The infinite product, of course, can be evaluated using e.g.~zeta function regularization,
as we do below.) 
The above formula can be interpreted as the result for the Witten 
index~\cite{Gautason:2024nru, PYiLectures}, which diverges as the volume of the 2d space.

\medskip

Our goal now is to regulate the IR divergence caused by the volume of the tangent space to 
the~$S^2$ transverse to the maximal giant in the calculation of the supersymmetric index. 
In the Hamiltonian analysis, the refinement is achieved by inserting the operator~$q^R$ in the 
trace~\eqref{eq:HSLanInd-R-trace}. 
In the functional integral formalism, this corresponds to twisting the fields around the time circle. 
Formally, one works in equivariant cohomology~\cite{Atiyah:1984px} 
(see~\cite{Cremonesi:2013twh,Pestun:2016qko,PYiLectures} for nice reviews). 
The twist implements the refinement as well as regulates the IR divergence by giving an 
effective mass to the flat directions, as we now proceed to discuss. 

Note that the charge~$R$ in the deformed theory commutes with the supercharge. 
The analogous statement in the Hamiltonian formalism is that the operator~$q^R$ 
preserves supersymmetry because the angular momentum in the Landau plane commutes 
with the supercharge~$Q$, as mentioned at the end of Section~\ref{sec:susyLan}. 
In particular, the R-charges of the boson-fermion pairs that cancel in the index are 
equal. Note that this is not the case for the R-charge of the full string theory---this subtlety 
was mentioned in the introduction to Section~\ref{sec:BPSbranes} as
the reason why the index is defined as a trace only over the~$\frac12$-BPS states. 
However, the fact that we are calculating an index allows us to deform the theory without changing 
the answer as long as the original boson-fermion cancellations are respected.

\medskip

In the Landau problem, which captures the fluctuations, we twist all the fields by the action of 
the global rotation symmetry~$L$ 
whose infinitesimal action is~$L \, x_i = -\varepsilon_{ij} \, x_j$,
~$L \, \lambda_i = -\varepsilon_{ij} \, \lambda_j$.\footnote{We use the same letter~$L$ for the 
AdS radius as well as the rotation; the usage should be clear from the context.}
Equivalently, we can make all the time derivatives covariant as~$\p_{t_E} \mapsto \p_{t_E} + \alpha L$. 
The deformation parameter~$\alpha$, which is the value of the external gauge field coupling 
to~$L$, is taken to be a complex number whose range we discuss below. 
The supersymmetry variations are also deformed as follows, 
\begin{equation}
\label{eq:EucSusyVars-alpha}
    \delta_\alpha \, x_i \= \lambda_i \,,  \quad 
    \delta_\alpha \, \lambda_i \= -  \, \dot{x}_i +  \alpha \, \varepsilon_{ij} \, x_j \,,
\end{equation}
with~$\varepsilon_{12} = -\varepsilon_{21}=1$.
The resulting algebra 
\begin{equation} \label{eq:twisted-algebra}
    \delta_\alpha^2 \= - \, \p_{t_E} - \alpha L \
\end{equation}
defines the equivariant cohomology with respect to~$L$.

One can check that the Lagrangian obtained by replacing the derivatives by covariant 
derivatives in~\eqref{eq:SLanEuc},
which we denote 
by~$\mathcal{L}^E_\text{SLan}(\alpha) $,  is invariant under the supersymmetry 
transformations~\eqref{eq:EucSusyVars-alpha}. 
In fact, this Lagrangian is actually~$\delta_\alpha$-exact. 
Indeed, following~\cite{Eleftheriou:2023jxr}, we define 
\begin{equation}
    V_1 \=-\frac{1}{2}\int_0^\gamma \, dt_E \, \bigl(\lambda_1 \,\delta_\alpha \lambda_1 
    + \lambda_2 \,\delta_\alpha \lambda_2 \bigr)\,, \qquad 
    V_2 \= i\, \int_0^\gamma \, dt_E \, \bigl(x_2 \, \delta_\alpha x_1 - x_1 \, \delta_\alpha x_2 \bigr)\,,
\end{equation}
and we have\footnote{Note that the action~$\delta_\alpha V_1$ equals the 
operator~$\frac12 B^2 \mathcal{R}^2=X_1^2+X_2^2$ 
introduced in~\eqref{eq:defR2}, for $B = i \alpha$.}  
\begin{equation} \label{eq:Lag-Euc-twisted}
\begin{split}
         \delta_\alpha \bigl( V_1+ b\, V_2 \bigr) \= &  \int_0^\gamma \, dt_E \left(-\frac{1}{2}\dot{x}_1^2 -\frac{1}{2}\dot{x}_2^2 
         +\frac{1}{2}\dot{\lambda}_1 \lambda_1 +\frac{1}{2}\dot{\lambda}_2\lambda_2 \right.
        \\ 
        &\left. - \left(\alpha- i b\right)(-\dot{x}_1 x_2 + \dot{x}_2 x_1) 
        - \frac{1}{2} \alpha(\alpha-2ib)(x_1^2 + x_2^2) + (\alpha-2ib) \lambda_1\lambda_2 \right)\,.
\end{split}
\end{equation}
Now it is easy to check that
\begin{equation} \label{eq:twistedaction}
     \mathcal{S}^E_\text{SLan}(\alpha)  \= \int_0^\gamma \, dt_E \, \mathcal{L}^E_\text{SLan}(\alpha) 
     \= \delta_\alpha \Bigl( V_1+ \frac{B}{2} \, V_2 \Bigr) \,.
\end{equation}

\medskip

Before computing the Euclidean path integral with  the twisted action~\eqref{eq:twistedaction}, 
we connect it to the trace interpretation discussed in the previous section. 
The canonical Hamiltonian corresponding to the  $\alpha$-deformed Lagrangian~\eqref{eq:Lag-Euc-twisted} 
takes the following simple form, 
\begin{equation}
    H_\text{SLan}(\alpha) \= \frac{1}{2}\, \bigl(p_1^2+p_2^2\bigr) 
    + \frac{1}{2}\left(\frac{B}{2}\right)^2\bigl(x_1^2+x_2^2\bigr)-\left(\frac{B}{2} 
    +i \alpha\right)\bigl(x_1 p_2-x_2 p_1\bigr) -\frac12 (\alpha-iB) \, [\lambda_1,\lambda_2]\,.
\end{equation}
This obeys 
\begin{equation} \label{eq:HdefH0rel}
    H_\text{SLan}(\alpha) \= H_\text{SLan}(0) -i\alpha \, \bigl(\, \widehat{L} + \tfrac12 \, \bigr) \,,
\end{equation}
where $H_\text{SLan}(0)$ and $ \widehat{L}$ are the untwisted Hamiltonian and the angular 
momentum operator, respectively, studied in Section \ref{sec:susyLan}. 
The twisted path integral thus computes the following trace,
\begin{equation} \label{eq:twistedtrace}
     \text{Tr} \, (-1)^F \, e^{-\gamma H_\text{SLan}\left( \alpha \right)} \= \text{Tr} \, (-1)^F \, 
     e^{-\gamma H_\text{SLan}\left(0 \right)} \, e^{i \alpha \gamma\widehat{L}} \,,
\end{equation}
where we have removed the zero-point shift in the angular momentum in accord with the 
physical vacuum condition of the brane as discussed in Section~\ref{sec:susyLan}. 
This shift will reappear at the end of the path integral analysis.

We can also derive the same result from the algebra~\eqref{eq:twisted-algebra}. 
In Lorentzian signature we have $-\partial_{t_E}= i \partial_t$. The rotation operator $L$ 
can be identified in the Lorentzian theory through 
its action on the bosonic fields, i.e., 
\begin{equation}
    -x_2 \partial_{x_1} + x_1 \partial_{x_2} = i(x_1p_2-x_2p_1) = i \widehat{L} \,,
\end{equation}
with a similar analysis holding in the fermionic sector.
The twisting $\partial_{t_E}\mapsto \partial_{t_E}+\alpha L$ thus leads precisely to~\eqref{eq:twistedtrace}.
As anticipated earlier, the path integral with the $\alpha$-twisted boundary conditions is 
equivalent to the Hamiltonian trace with an operator insertion~$q^{-\widehat{L}}$ 
with~$q=e^{-i \gamma \alpha}$ corresponding to the refinement by the R-charge 
as in~\eqref{eq:HSLanInd-R-trace} after including the ground state contribution of~$q^N$.

\medskip

Now we turn to the functional integral. The~$Q$-exact term~$\delta_\alpha(V_1 + b V_2)$ 
for~$\alpha = a + ib$,~$a, b \in \mathbb{R}$, has the following bosonic part,
\begin{equation} \label{eq:V1bV2def}
    \delta_\alpha \bigl( V_1+ b \, V_2 \bigr) 
    \biggr\vert^{\text{bos.}}_{\alpha = a + ib} 
    \= - \int_0^\gamma \, dt_E \left(\frac{1}{2}\left( \dot{x}_1 - a x_2 \right)^2 
    + \frac{1}{2}\left( \dot{x}_2 + a x_1 \right)^2 + \frac{b^2}{2}\left( x_1^2+x_2^2 \right)\right) \,,
\end{equation}
which is manifestly negative semi-definite. Therefore, the introduction of  $ \delta_\alpha \bigl( V_1+ b \, V_2 \bigr) 
    \bigr\vert_{\alpha = a + ib}$  in the action with a coefficient $t\geq 0$ does not change the 
    result of the path integral. 
In the limit~$t \to \infty$, the functional integral is dominated by its critical points. 
For~$b\neq 0$, the quadratic term in~$x_1,x_2$ forces the fixed points to be~$x_1 = x_2 = 0$. 
Thus, we have found a~$\delta_\alpha$-exact deformation which has the effect of 
localizing the path integral as well as lifting the flat directions. 

For~$b = 0$, the fixed points are  given by harmonic motion for~$x_i$ with frequency~$a$. 
The only solution respecting the periodicity of the Euclidean time circle is again~$x_i=0$ as long 
as the parameter~$a \gamma$ does not coincide with a 
multiple of~$2 \pi$. 
For~$a \gamma \notin 2\pi\mathbb{Z}$, we can continuously change the value of~$b$ from 
positive to negative values,
all the while keeping the action negative semi-definite so that the path integral is well-defined. 
In this sense, the path integral with the~$\delta_\alpha$-exact deformation provides a derivation 
of the analytic continuation for the GGE of the~$\frac{1}{2}$-BPS index.

\bigskip

The value of the twisted action is zero at the fixed point and, therefore, the contribution of the 
fixed point to the functional integral is given by 
the super-determinant of~$\delta_\alpha V$ with~$V=(V_1+bV_2)$, $\alpha=a+ib$, $a, b \in \mathbb{R}$. 
The calculation of the super-determinant is similar to the one given above for the undeformed theory. 
As before, we focus on the fluctuation determinant, and keep the overall normalization to be fixed at the end. 
The non-zero mode determinants are now given by the following expressions, 
\be 
\begin{split}
\text{Bos}: & \qquad \prod_{n=1}^\infty \, \text{det}  
\begin{pmatrix} 
4 \pi^2 n^2 + |\alpha|^2 \gamma^2 & 4 \pi i a \gamma n \\
- 4 \pi i a\gamma n & 4 \pi^2 n^2 + |\alpha|^2 \gamma^2  
\end{pmatrix}^{-1} \= 
\prod_{n=1}^\infty \,  \bigl| 4 \pi^2 n^2 - \alpha^2 \gamma^2 \bigr|^{-2} \,, \\
\text{Fer}: & \qquad \prod_{n=1}^\infty \, \text{det}  
\begin{pmatrix} 
2 \pi n &  -i \overline{\alpha} \gamma \\
i \overline{\alpha} \gamma & 2 \pi n  \\
\end{pmatrix} \= 
\prod_{n=1}^\infty \,  \bigl( 4 \pi^2 n^2 - \overline{\alpha}^2 \gamma^2 \bigr) \,.
\end{split}
\ee 
As in the undeformed theory, the non-zero modes of the fermions cancel the non-zero modes of one boson, 
but now the modes of the unpaired boson have an additional mass term proportional to~$\alpha^2$. 
Note that the~$\alpha$-dependence of the paired bosons and fermions is antiholomorphic, 
and that of the unpaired boson is holomorphic. 
The bosonic zero mode is now lifted by the quadratic terms in~\eqref{eq:V1bV2def}. The 
corresponding Gaussian integral over~$x^i_0$ produces a factor of~$1/\alpha^2 \gamma^2$, 
replacing the volume divergence in the undeformed theory. 
The integral over the fermionic zero modes produces a factor of~$i \alpha \gamma$. 
Upon putting together the various pieces, we obtain
\be \label{eq:SdetResult}
\begin{split}
\text{SDet}(\delta_\alpha V)^{-1} & \= 
\mathcal{N} \, \frac{1}{\alpha^2 \gamma^2} \, (i\alpha \gamma) 
\prod_{n=1}^\infty  \frac{1}{4 \pi^2 n^2} \frac{1}{\bigl( 1- \alpha^2 \gamma^2 / 4 \pi^2 n^2 \bigr)} \\
& \= \mathcal{N} \frac{1}{\alpha \gamma} \, \frac{i\alpha \gamma/2}{\sin(\alpha \gamma/2)} 
\= \frac{\mathcal{N}}{2i\sin(-\alpha \gamma/2)} \,.
\end{split}
\ee 
In going to the second line we use the zeta function regularization and the infinite product 
representation of the sine function.

As mentioned above, the measure of the integrals are chosen such that there is 
an exact cancellation between one boson and the two fermions. 
The remaining infinite product corresponds to exactly the determinant of a free boson, with 
the normalization constant~$\mathcal{N}$ being the product of the corresponding free 
Gaussian integrals. 
If we are interested in the partition function of the free boson, this would be set to one, 
as corresponds to the ultralocal measure. 
The only non-trivial role of $\mathcal{N}$ in the present treatment is to account for the 
zero-point angular momentum~\eqref{eq:HdefH0rel} when we 
relate the fluctuation determinant to the brane problem. 
This fixes the normalization to be~$\mathcal{N}=e^{-i \gamma \alpha/2}$, so that the final 
result is, with~$q=e^{-i\gamma \alpha}$,
\be \label{eq:SDetval}
\text{SDet}(\delta_\alpha V)^{-1} \=
\frac{1}{1-q^{-1}} \= -\frac{q}{1-q} \,.
\ee
To compare this expression with the full index~\eqref{eq:INtrace1}, we need to 
grade the states by~$R = - i\partial_{\phi} = N-\widehat{L}$, which leads to an additional 
contribution of~$q^N$ as discussed above~\eqref{eq:HSLanInd-R-trace}. 

As in the Hamiltonian analysis, only one of the two expansions~\eqref{eq:SDetval} 
in~$q$ or~$q^{-1}$ is convergent, depending on the sign of~$\text{Im}(\alpha)$. 
The two expressions are interpreted as the expansions on either side of the wall, 
with the first expansion in~$q^{-1}$ capturing the physical fluctuations and the second 
expansion in~$q$ capturing the GGE.\footnote{These 
expressions were also calculated recently in~\cite{Lee:2024hef} by 
analyzing the physical partition function of the single unpaired boson around an unstable 
saddle using Lefshetz thimbles.}  
Note that, in the Hamiltonian formalism, $\text{Tr}_\text{LLL}\, (-1)^F e^{-i\gamma (a+ib) \widehat{L}}$ 
is convergent for~$a \in \mathbb{R}$ for both signs of $b$, as consistent with the 
wall-crossing phenomenon upon identifying~$b=B/2$.

\bigskip

We end this section with a few comments. Firstly, the  deformation parameter~$\alpha$ 
plays two roles---firstly, it couples to the angular 
momentum and thus refines the trace through the parameter~$q=e^{- i \gamma \alpha}$, 
and, secondly,~$\text{Im}(\alpha)$ is identified with the magnetic field on the brane. 
The functional integral is well-defined for generic values of~$\alpha \in \mathbb{C}$, and this 
allows us to effectively change the sign of the magnetic field 
to obtain the convergent GGE, 
through an analytic continuation in~$\alpha$ without crossing the singular point~$q=1$. 
In this sense, it provides a more rigorous basis to the wall-crossing phenomenon seen in the Hamiltonian analysis. 

Secondly, the answer for the index of the~$U(1)$-character valued index of the type we consider 
can also be deduced from a fixed point formula~\cite{Goodman:1986wq,Witten:1983ux,Atiyah:1984px}.

Finally, it is satisfying to explain the analytic continuation of the giant-graviton index as an instance 
of wall-crossing. 
On a broader note, the elements involved in the above analysis of the wall-crossing of the giant 
gravitons---an infrared divergence that 
is cured by introducing a supersymmetric regulator, states moving in and out of the spectrum 
due to their normalizability 
properties, different expansions of analytic functions capturing the jumps in the spectrum---are 
precisely the ones involved in diverse wall-crossing phenomena. 
Some examples are the index of the Dirac operator in Taub-NUT space~\cite{Pope:1978zx}, 
the monopole moduli space \cite{Gauntlett:1999vc}, 
as well as supersymmetric black holes~\cite{Dabholkar:2007vk,Dabholkar:2012nd}.
It would be nice to give similar explanations for the richer structure of analytic continuations in 
GGEs with less supersymmetry~\cite{Bourdier:2015wda,  Imamura:2022aua}.

\section{Boundary indices for symplectic and orthogonal gauge groups} 
\label{sec:boundary-indices}

In this section we review some elementary aspects of $\mathcal{N}=4$ SYM with $SO$ 
and $Sp$ gauge groups in string theory.  
We explain how these theories arise from the $U(N)$ $\mathcal{N}=4$ SYM worldvolume 
theory of a stack of $N$ D3-branes 
after placing an orientifold plane parallel to them, with different projections leading to 
different gauge groups. For each resulting gauge group, we enumerate 
the $\frac{1}{2}$-BPS gauge invariant operators and 
derive the corresponding index.
Counting arguments arriving at these results appear in~\cite{Caputa:2013hr,Caputa:2013vla}. 
A related study on operators dual to giant gravitons for the orthogonal gauge group theory 
can be found in~\cite{Lewis-Brown:2018dje}. In this section we follow the more recent 
treatment of~\cite{Arai:2018utu}. 
After deriving the expressions for the $\frac{1}{2}$-BPS index for each gauge group, 
we proceed to also write it mathematically in GGE form.

\subsection{Orientifolds in flat space}

$\CN=4$ SYM theory with~$SO$ or~$Sp$ gauge groups can be constructed in string 
theory as follows. We start with a stack of $N$ D3-branes in 10-dimensional flat space. 
We gauge the theory by the orientifold symmetry that is a combination of the involution~$\beta$ 
acting as a reflection of the 6 orthogonal directions to the D3-branes 
and the world-sheet parity operator~$\Omega$ that acts by reversing the orientation of the 
strings~\cite{Evans_1997, Dabholkar:1997zd, Polchinski:1998rr, Dai:1989ua, Gimon:1996rq, 
Polchinski:1996fm,Polchinski:1996na}. 
These symmetries satisfy
$\Omega \beta = \beta \Omega$, $\beta^2=\Omega^2=1$, and therefore the  orientifold 
symmetry~$\Omega \beta$ has a~$\mathbb{Z}_2$ action.

The fixed points of the orientifold action constitute a $3+1$ dimensional plane parallel to 
the D3-branes, called an orientifold O3-plane. 
While the shape of D-branes fluctuate dynamically due to the fluctuations of open strings 
ending on them,  
the O-plane by  definition is independent of string boundary conditions and is  
not dynamical. 
Moreover this O3-plane is charged under the same RR $4$-form gauge potential 
as the parallel D3-branes, and it also breaks the same half of the supersymmetries, 
leading to a gauge theory with 16 supercharges.

To determine the gauge group of this orientifold theory, we need to consider what 
happens to the Chan-Paton factors, 
namely the wavefunctions represented by~$N\times N$ matrices $\lambda_{ij}$, 
where~$i$ and~$j$ label the D-branes on which the endpoints of an open string lie, 
with orientation from~$i$ to~$j$.
The world-sheet parity operator~$\Omega$ acts as~$\sigma \rightarrow \pi - \sigma$, 
where $\sigma$ is the spacelike coordinate on the world-sheet. By requiring that the action 
of~$\Omega$ is a symmetry of general open string amplitudes, one obtains that it has 
the following general form on Chan-Paton factors, 
\begin{equation}
    \Omega \lambda \= M \lambda^T M^{-1} \,.
\end{equation}
Since~$\Omega$ is an involution, we have 
\begin{equation}
    \lambda \= \Omega^2 \lambda \= MM^{-T}\lambda M^{T}M^{-1} \,,
\end{equation}
and, by Schur's lemma, $MM^{-T}$ is proportional to the identity. Up to a choice of basis, 
this leads to two cases, the symmetric one,
\begin{equation}
    M\=M^T\=I_N,
\end{equation}
and the antisymmetric one
\begin{equation} \label{eq:Msymp}
    M\=-M^T\= i J \,,
    \qquad J \= \begin{pmatrix}
0 & I_{N/2} \\
-I_{N/2} & 0 
\end{pmatrix}.
\end{equation}
Note that the antisymmetric case can only occur whenever $N$ is even.
Requiring that photons survive the~$\Omega$-projection means 
that~$\Omega \lambda = -\lambda$~\cite{Polchinski:1996fm, Polchinski:1996na}.
In the symmetric case, this leads to~$\lambda=-\lambda^T$, and thus the gauge group is $SO(N)$. 
In the antisymmetric case, we have~$\lambda = -M \lambda^T M = J \lambda^T J$ 
with~$M = iJ$ and~$J$ being the standard symplectic form as in~\eqref{eq:Msymp}. 
Thus we see that the gauge group is~$Sp(k)$, where~$k=N/2$.

These two cases also correspond to a difference in the sign of the RR charge of the O3-plane. 
If we denote the RR charges of the O3-plane and a D3-brane as $Q_{O3}$ and $Q_{D3}$ 
respectively, they satisfy the following relation  
\begin{equation} \label{eq:o3-plane-charge}
    Q_{O3} \= \pm 2^{-1} Q_{D3} \,.
\end{equation}
The symmetric or antisymmetric projections of the $N \times N$ matrices of the gauge fields and 
the scalars analysed above correspond to either a positive or a negative sign, respectively, 
in the above expression.

\subsection{Counting and GGEs}

For free~$\mathcal{N}=4$ SYM, the~$\frac{1}{2}$-BPS gauge invariant operators, corresponding 
to a choice of supercharge, are holomorphic functions of one of the matrix-valued complex scalars 
of the theory, say $X$, which transforms in the adjoint representation of the gauge group.
The ring of such functions is freely generated by a finite list of operators. 
The number of independent operators of this type is given by the rank of the gauge 
group~$G$ of the theory.

We proceed to determine this list of generators for various gauge groups.
Firstly, note that the Cayley-Hamilton theorem implies that for a gauge group of rank~$N$ we 
can write any trace~$\text{Tr}(X^i)$ with $i>N$ as a polynomial combination of traces 
with~$i \leq N$, and so such traces are not part of the list of generators. In fact, we know 
from the results of~\cite{PROCESI1976306} that any polynomial relation among traces of complex 
matrices is a consequence of the Cayley-Hamilton theorem.

\medskip

\subsection*{$\bullet$ $G=U(N)$\, \&  $G=SU(N)$}

For $G=U(N)$, the field $X$ is an element of the complexification of the Lie algebra~$u(N)$, i.e.~it is an arbitrary complex matrix. 
The fact that the Cayley-Hamilton theorem is the only constraint among the traces  
implies that the ring of gauge invariant operators is freely generated by the following generators, 
\begin{equation} \label{eq:UNgens}
    \text{Tr}(X)\,, \, \text{Tr}(X^2) \,, \, \ldots \,, \, \text{Tr}(X^N) \,.
\end{equation}
Consequently, we obtain the following index, 
\begin{equation}
    I_{U(N)}(q) \= \frac{1}{(q)_N}\,,
\end{equation}
which can be written in GGE form as follows 
\begin{equation}
\label{GGE_UNold}
    {I_{U(N)}}(q)  \= \frac{1}{(q)_\infty}\,\sum_{m=0}^\infty (-1)^m \, \frac{q^{\binom{m+1}{2}}}{(q)_m}q^{mN} \, .
\end{equation}
\medskip

\ndt For $G=SU(N)$, we have the same list of generators~\eqref{eq:UNgens}, but 
without~$\text{Tr}(X)$ since~\hbox{$\text{Tr}(X)=0$}. Upon removing this generator, we obtain the index
\begin{equation}
    I_{SU(N)}(q) \= (1-q) \, I_{U(N)}(q) \= \frac{(1-q)}{(q)_N} \= \frac{1}{(q^2;q)_N}\,,
\end{equation}
and the GGE 
\begin{equation} \label{eq:GGE-SUN}
    {I_{SU(N)}}(q) \= \frac{1}{(q^2;q)_\infty}\,
    \sum_{m=0}^\infty (-1)^m \,\frac{q^{\binom{m+1}{2}}}{(q)_m}q^{mN}\, .
\end{equation}
We see that the difference between the $U(N)$ and $SU(N)$ groups in the giant graviton 
expansion is entirely due to the supergravity modes.

\medskip

\subsection*{$\bullet$ $G= SO(N)$}

For $G=SO(N)$, the field $X$ is an element of the complexification of the Lie algebra~$so(N)$, and 
is a complex antisymmetric matrix. 
This implies that traces of odd powers of $X$, which are also antisymmetric, vanish. 
The set of generators of $\frac{1}{2}$-BPS gauge invariant operators changes depending 
on whether~$N$ is even or odd. In both cases, we have traces of even powers of~$X$, and 
for even~$N$ there is an additional generator. 

When~$N=2k+1$, the ring of gauge invariant operators is freely generated by 
\begin{equation}\label{SO_odd}
    \text{Tr}(X^2) \,, \, \text{Tr}(X^4)\,,\, \ldots, \, \text{Tr}(X^{2k}) \,,
\end{equation}
which leads to the following index, 
\begin{equation}
    {I_{SO(2k+1)}}(q) \= \frac{1}{(q^2)_k}\,. 
\end{equation}
The GGE form of the index then follows from the GGE of the~$U(N)$ theory \eqref{GGE_UNold} 
with the replacement~$q \mapsto q^2$
\begin{equation}\label{SO_odd_GGE}
    {I_{SO(2k+1)}}(q)\= \frac{1}{(q^2)_\infty}\, \sum_{m=0}^\infty \, (-1)^m \, 
    \frac{q^{2\binom{m+1}{2}}}{(q^2)_m} \, q^{2mk}\, .
\end{equation}

\medskip

When~$N=2k$, in addition to the traces of even powers, we have a new gauge-invariant operator,
namely the Pfaffian of the matrix~Pf($X$). (The Pfaffian of an antisymmetric matrix of size~$N$ vanishes for odd~$N$.) 
The Pfaffian is a degree $k$ polynomial in the entries of~$X$, with integer coefficients that are 
combinatorial factors that depend on the size of the matrix.
Using the Cayley-Hamilton theorem, we can write $\text{Tr}(X^{2k})$ as a polynomial in the traces 
of smaller powers of~$X$ and~$\text{det}(X)$. Since we have $\text{det}(X)=\text{Pf}(X)^2$, 
$\text{Tr}(X^{2k})$ can be written as a polynomial in the traces of smaller powers of $X$ and Pf($X$). 
This means that in this case the list of free generators is
\begin{equation}
    \text{Tr}(X^2)\,, \, \text{Tr}(X^4)\,, \ldots, \, \text{Tr}(X^{2k-2})\,, \, \text{Pf}(X)\,.
\end{equation}
The inclusion of $\text{Pf}(X)$ and the removal of $\text{Tr}(X^{2k})$ from the list of generators 
corresponds to multiplying $I_{SO(2k+1)}$ by $(1-q^{2k})/(1-q^k)$, leading to 
\begin{equation}
    {I_{SO(2k)}}(q)  \= \frac{1-q^{2k}}{1-q^k} \,{I_{SO(2k+1)}}(q) 
    \= (1+q^k) \,{I_{SO(2k+1)}}(q) \= \frac{1+q^k}{(q^2)_k}\, .
\end{equation}
Finally, using~\eqref{SO_odd_GGE}, we obtain the GGE form,
\begin{equation} \label{eq:gge-so2k}
   {I_{SO(2k)}}(q)  \= \frac{1}{(q^2)_\infty}\,\sum_{m=0}^\infty (-1)^m \,
   \frac{q^{2\binom{m+1}{2}}}{(q^2)_m}\,(1+q^k)\,q^{2mk}\, .
\end{equation}

\medskip

\subsection*{$\bullet$ $G=Sp(k)$}

For $G=Sp(k)$, the field $X$ is an element of the complexification of the Lie algebra~$sp(k)$, and satisfies
\begin{equation}
    X \= J X^T J \, ,   
\end{equation}
where $J = \begin{pmatrix}
0 & I_k \\
-I_k & 0 
\end{pmatrix} $ is the standard symplectic form matrix that satisfies $J^{-1}=-J$.
This implies that traces of odd powers of $X$ vanish, leading to the following list of 
free generators, 
\begin{equation}
    \text{Tr}(X^2) \,, \, \text{Tr}(X^4)\,, \, \ldots, \, \text{Tr}(X^{2k}) \,.
\end{equation}
Note that the Pfaffian operator is not well-defined for a general element in the Lie algebra of $Sp(k)$, 
and therefore the Pfaffian doesn't appear at all in the list of operators for $Sp(k)$.
Since the lists of free generators for~$Sp(k)$ and for~$SO(2k+1)$ are exactly the same, so are the indices, i.e.,
\begin{equation}
    {I_{Sp(k)}}(q) \= {I_{SO(2k+1)}}(q) \= \frac{1}{(q^2)_{k}}\,, 
\end{equation}
and the corresponding GGEs are also the same, 
\begin{equation} \label{eq:sp2k-so2k1-gge}
     {I_{Sp(k)}}(q)\={I_{SO(2k+1)}}(q)\= \frac{1}{(q^2)_\infty}\,\sum_{m=0}^\infty (-1)^m \,
     \frac{q^{2\binom{m+1}{2}}}{(q^2)_m}q^{2mk}\, .
\end{equation}
Recall that the result ${I_{Sp(k)}}(q)={I_{SO(2k+1)}}(q)$ is to be expected since 
the~$\mathcal{N}=4$ theories with $SO(2k+1)$ and $Sp(k)$ gauge groups are 
duals~\cite{MONTONEN1977117,  Witten:1998xy}, and that the theory 
with~$SO(2k)$ gauge group is self-dual.

\section{\texorpdfstring{$\frac12$}{1/2}-BPS branes and their fluctuations in the 
orientifold theories\label{sec:BPSbranesOrient}}

The BPS locus of the orientifolded theory can be split into twisted and untwisted sectors. 
The untwisted sector consists of the projection of the BPS locus of the parent theory, 
while the twisted sector has no counterpart in the original theory.

\subsection{Orientifolded bulk duals}

As explained in \cite{Witten:1998xy}, the holographic dual to $\mathcal{N}=4$ SYM 
in four dimensions with gauge group~$SO(N)$ or~$Sp(k)$ is type IIB string theory on 
an AdS$_5 \times \mathbb{RP}^5$ orientifold. The duality is obtained by placing~$N$~D3-branes 
in flat ten-dimensional spacetime at an orientifold threeplane. The orientifold action replaces 
the transverse space~$\mathbb{R}^6$ by~$\mathbb{R}^6/\mathbb{Z}_2$ and the~$S^5$ 
in the near-horizon geometry by~$\mathbb{RP}^5 =S^5/\mathbb{Z}_2$. 
The~$\mathbb{Z}_2$ action identifies antipodal points~\hbox{$x\sim -x$} for~$x\in S^5$, reversing the 
orientation of strings as they go around a non-contractible cycle in $\mathbb{RP}^5$. 

The particular gauge group of the dual theory depends on the value of the discrete torsion of the 
two-form fields~$B_{NS\, NS}$ and~$B_{RR}$. More precisely, after orientifolding the two-form 
fields~$B$ become twisted two-forms, meaning that their field strength~$H=dB$ determines a 
cohomology class~$[H]$ that takes values 
in~$H^3(\mathrm{AdS}_5 \times \mathbb{RP}^5, \widetilde{\mathbb{Z}}) = \mathbb{Z}_2$, 
where~$\widetilde{\mathbb{Z}}$ are twisted coefficients which change sign along a non-contractible loop. 
Thus, there are in total four possible options of discrete torsion for the two~$B$ fields. 
The boundary gauge group is identified by the action of the~$SL(2,\mathbb{Z})$ on the~$B$ fields. 
When there is no torsion the group is~$SO(2k)$, while for non-zero torsion the groups are~$SO(2k+1)$ 
or~$Sp(k)$. More details can be found in \cite{Witten:1998xy}. 
A relevant point is that a threebrane can wrap the submanifold~$\mathbb{RP}^3 \subset \mathbb{RP}^5$ 
only if there is no torsion. This is consistent with the fact that only for the~$SO(2k)$ gauge group 
is there a Pfaffian operator.

The orientifolded theory has no extra winding sector from strings wrapping a non-contractible 
loop in~$\mathbb{RP}^5$ since the orientation must be reversed when going around the loop. 
Furthermore, there are no extra open string sectors because the orientifold has no fixed points. 
The spectrum of the theory at weak coupling is obtained from the states which are invariant 
under the orientifold projection in AdS$_5\times S^5$.

\subsection{Orientifold projection on the giants}

To see the action of the antipodal identification on~$S^5$, we  describe it as the 
subset of~$\mathbb{R}^6$ defined by $x_1^2+\dots +x_6^2 = L^2$~with the following coordinates, 
\begin{equation}
\begin{split}
          & x_1 \= L\cos \theta \sin\phi \,, \;\,\;\qquad\qquad\quad\qquad x_2 \= L\cos\theta\cos\phi \,, \\
         & x_3 \= L\sin\theta \cos\chi_1\,, \qquad\qquad\qquad\quad\; 
         x_4 \= L\sin \theta \sin\chi_1\cos\chi_2 \,, \\
        &   x_5 \= L\sin \theta \sin\chi_1\sin\chi_2\cos\chi_3 \,, \qquad x_6 \= L\sin \theta \sin\chi_1\sin\chi_2\sin\chi_3\,.
\end{split}
\end{equation}
The antipodal map~$x_i\to -x_i$~for the~$\mathbb{R}^6$ coordinates is equivalent to
\begin{equation} \label{eq:involution}
    \theta\,\to\, \theta\, , \qquad \phi \,\to\, \phi + \pi\,, \qquad \chi_1 \,\to\, \pi -\chi_1\,, 
    \qquad \chi_2 \,\to\, \pi-\chi_2\, , \qquad \chi_3 \to \chi_3+\pi \, ,
\end{equation}
in the~$S^5$ coordinates. On the giants, the involution~\eqref{eq:involution} acts on 
the worldvolume coordinates~$\chi_i$ as well as on the center of mass position~$(\theta,\phi)$.

\begin{figure}[ht]
    \centering
    \begin{tikzpicture}[scale=0.9, every node/.style={transform shape}]

\draw[thick] (-4,0) arc[start angle=180, end angle=360, x radius=4cm, y radius=1cm]; 

\draw[dashed] (4,0) arc[start angle=0, end angle=180, x radius=4cm, y radius=1cm];

\draw[thick] (-4,0) arc[start angle=180, end angle=0, radius=4cm]; 

\draw[thick] (-3.45,2) arc[start angle=180, end angle=360, x radius=3.45cm, y radius=0.5cm]; 

\draw[dashed] (3.45,2) arc[start angle=0, end angle=180, x radius=3.45cm, y radius=0.5cm]; 

\node at (0,4.24) {$\theta=\frac{\pi}{2}$};

\node at (4.5,0) {$\theta=0$};

\draw[thick] (4,2.2) arc[start angle=360, end angle=0, radius=0.3cm];

\draw[thick] (-3.4,2.2) arc[start angle=360, end angle=0, radius=0.3cm];

\draw[->] (-0.3,1.7) to[out=-40, in=-140] (0.3,1.7);

\draw[<-] (-0.3,2.65) to[out=40, in=140] (0.3,2.65);

\draw[thick] (-4,2.2) arc[start angle=180, end angle=360, x radius=0.3cm, y radius=0.1cm]; 

\draw[dashed] (-3.4,2.2) arc[start angle=0, end angle=180, x radius=0.3cm, y radius=0.1cm];

\draw[thick] (3.4,2.2) arc[start angle=180, end angle=360, x radius=0.3cm, y radius=0.1cm]; 

\draw[dashed] (4,2.2) arc[start angle=0, end angle=180, x radius=0.3cm, y radius=0.1cm];

\end{tikzpicture}
\hspace{10mm}
\begin{tikzpicture}[scale=0.9, every node/.style={transform shape}]

\draw (0,0) circle (3cm);

\fill (0,0) circle (0.03cm);

\def\radius{3cm}
\def\sphereradius{0.3cm} 
\def\angleone{45} 
\def\angletwo{225} 

\foreach \angle in {\angleone, \angletwo} {
    \fill (\angle:\radius) circle (\sphereradius);
}

\foreach \angle in {135, 315} {
    \draw[->, thick, black] (\angle:\radius - \sphereradius) arc (\angle:\angle + 45:\radius - \sphereradius);
}

\end{tikzpicture}
    \caption{Left: Configuration of a pair of D-branes wrapping an $S^3\subset S^5$ and rotating 
    around the $\phi$ circle at opposite points for a fixed value of $0<\theta<\frac{\pi}{2}$. 
    Right: Azimuthal view of the same configuration.}
    \label{fig:enter-label}
\end{figure}
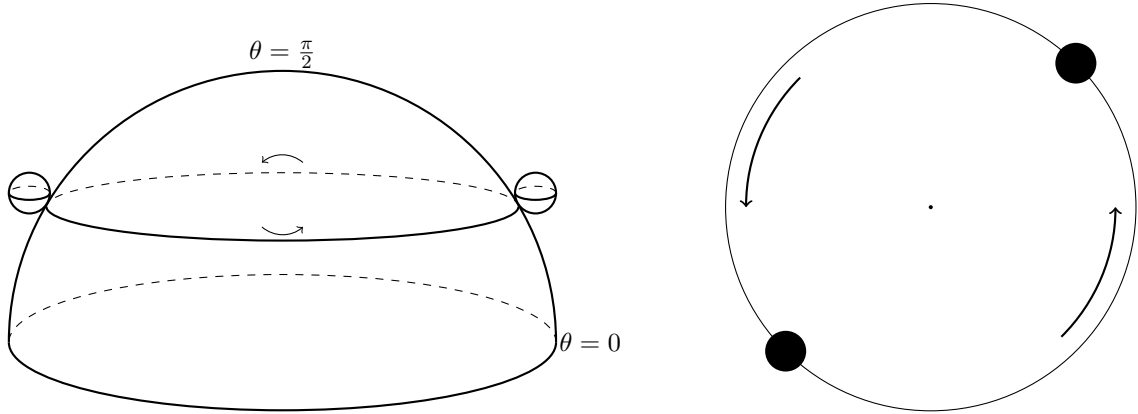

For generic~$\theta_0<\frac{\pi}{2}$, the configurations in AdS$_5\times S^5$ that are invariant under the 
above involution are pairs of giants circling around opposite points in the circle parametrized by~$\phi$  
(as described in~\cite{Mukhi:2005cv} in the context of bubbling geometries) with their antipodal points 
identified. In the limiting case of maximal size, as~$\theta_0$ approaches~$\frac{\pi}{2}$, the two giants 
become coincident. 
The set of all these pairs describes the untwisted~$\frac{1}{2}$-BPS moduli space. In addition to the 
antipodal identification of the worldvolumes of pairs of giants, the orientation of the open string 
excitations on the branes gets reversed under orientifolding. We describe its effect on the 
spectrum of excitations below.

\subsection*{BPS fluctuations of maximal giants on $\mathbb{R}\mathbb{P}^5$}

We turn to the analysis of the fluctuations of maximal giants in AdS$_5\times \mathbb{RP}^5$. 
The brane configurations in AdS$_5\times \mathbb{RP}^5$ can be described as those in 
AdS$_5\times S^5$ that are invariant under the orientifold projection. 
This can be achieved by taking a pair of branes as in~Figure~\ref{fig:enter-label}. 
In this section we analyze this type of brane.
In addition, there is the possibility of an 
isolated brane sitting at the fixed point which is 
also invariant under the geometric involution---but this is only possible when 
there is no discrete torsion, which implies that we have an~$SO(2k)$ theory~\cite{Witten:1998xy}. 
We analyze this type of brane in Section \ref{sec:rigidbranes}.

As explained earlier, in the untwisted sector there is no single maximal giant graviton. 
The first contribution to the giant graviton expansion comes from a pair of maximal giants 
in the covering space 
where the antipodal points of their worldvolumes are identified and the orientation of the 
open strings on the maximal giants which form a non-contractible cycle gets reversed under orientifolding. 
As in the boundary, there are two possible orientifold projections for the open string excitations 
on the coinciding brane worldvolumes. In this section we assume that the matrices describing 
the bosonic fields in the worldvolume theory of the coincident giants are antisymmetric 
as well as hermitian, i.e., imaginary and antisymmetric and therefore the gauge group 
on the pair of coincident branes is~$SO(2)$. We deal with the other orientifold projection in the next section.

The Lagrangian for the two coincident maximal giants is obtained by promoting the 
variables~$x,y$ in~\eqref{eq:Lag-landau-w-flux} to matrix variables~$X,Y$ which are imaginary and 
antisymmetric, and then taking the trace. 
Further, recall that for AdS$_5\times S^5$ the prefactor~$\frac{N}{L}$ in the Lagrangian appears 
after integrating the angular variables in \eqref{eq:action-d3} from the volume of the three-sphere 
wrapped by the D3-brane and the quantization condition for the 
AdS radius~$L\sim N^{1/4}$.\footnote{The parameter $N$ appears in the covering space 
through the quantization of the~$5$-form flux over~$S^5$ \eqref{eq:5-form-flux-N}, 
which fixes the relation between the AdS radius $L$ and~$N$ to be $L^4 =4\pi g_s \alpha'^2 N$. 
In particular, it implies the relation~$\mathrm{vol}(S^3) T_3 L^4 = N$ between the D-brane 
tension appearing in the action and $N$.}  
The orientifold projects the~$S^5$ to $\mathbb{RP}^5$  and the~$S^3$ to~$\mathbb{RP}^3$, 
thus reducing the volume by two. It therefore effectively reduces the total flux over~$\mathbb{RP}^5$ 
to be half the original value.\footnote{Here there is a subtlety, namely that there seems
to be an additional shift of~$\pm \frac14$ in the flux due to the orientifold plane 
(see e.g.~\cite{Bergman:2001rp,Giombi:2020kvo}). In~\cite{Tachikawa:2018njr} it is shown that this 
fractional flux is effectively cancelled by an anomaly in the worldvolume fermionic theory. 
We interpret this in our system to mean that the  supersymmetric Lagrangian of the probe branes 
has half-integer flux, as in the classical theory. 
}

The Lagrangian for two 
coincident maximal giants in the orientifold theory is thus
\begin{equation} \label{eq:Lag-landau-cartesian-2-w-flux}
     \mathcal{L}^{(2)}_\text{Lan} \= \frac{N}{2L} \mathrm{Tr}\biggl( 
    \frac{1}{2} \, \bigl(\dot{X}^2+  \dot{Y}^2 \bigr) + \frac{1}{L}(X\dot Y - Y\dot X) 
    -L(X\dot{Y}-Y\dot{X})(X^2+Y^2)^{-1}  \biggr) \,,
\end{equation}
where $X,Y$ are $2\times 2$ hermitian and antisymmetric matrices.  
Since the Lagrangian is quadratic, and the product of two $2\times 2$ antisymmetric matrices 
is proportional to the identity, 
the matrix Lagrangian becomes the original Landau problem \eqref{eq:Lag-landau-cartesian-2-w-flux} 
times the $2\times 2$ identity matrix. 
Taking the trace gives an overall factor of 2 so that writing the upper-right entries of $X,Y$ respectively 
as $ix, iy$ with $x,y$ real variables, 
the Lagrangian \eqref{eq:Lag-landau-cartesian-2-w-flux} becomes precisely the previous 
Lagrangian \eqref{eq:Lag-landau-w-flux},
\begin{equation} \label{eq:Lag-landau-cartesian2}
     \mathcal{L}^{(2)}_\text{Lan} \= \frac{N}{L} \biggl( 
    \frac{1}{2} \, \bigl(\dot{x}^2+  \dot{y}^2 \bigr) + \frac{1}{L}(x\dot y - y\dot x) \Bigl(
    1-\frac{L^2}{x^2+y^2}    \Bigr)    \biggr) \,.
\end{equation}
Again, discarding the solenoid term, which  shifts the angular momentum by $N$, the Hamiltonian 
(up to rescaling by factors of $N/L$) takes the same form as \eqref{eq:Hamiltonian-Landau-rescaled},
\begin{equation}\label{eq:HamNoFlux2}
     \mathcal{H}^{(2)}_\text{Lan}   \=  \frac{L}{2N} \bigl(p_x^2+p_y^2 \bigr) \,+\, 
     \frac{N}{2L}\frac{1}{L^2} \bigl(x^2 + y^2 \bigr) 
     \,-\, \frac{1}{L} \bigl(x p_y-yp_x \bigr) \,.
\end{equation}
When written in terms of the creation and annihilation operators there is no modification with 
respect to the treatment in covering space, and it reads exactly as in \eqref{eq:Ham-creation-ann}. 
The bosonic angular momentum operator takes the following form
\begin{equation} \label{eq:L-angular-2}
    \widehat{L}_{(2)} \= \frac{1}{2}\mathrm{Tr} (X P_Y - Y P_X) \=  xp_y-yp_x  \,,
\end{equation}
where the factor $\frac{1}{2}$ is introduced because we are considering two giant gravitons in the 
covering space which are identified in the projection. 
The operator $\widehat{L}_{(2)}$ remain unchanged with respect to the unorientifolded theory. 
In particular, its eigenvalues~$\ell$ take integer values as before.

The derivation of the index on the space of fluctuations follows the same principles as in 
the~$U(N)$ case. The only difference arises from the global identification due to the orientifold. 
This is true on both sides of the wall.  Upon reintroducing the solenoid term for the two coincident 
maximal giants, the spectrum of the R-charge of physical brane fluctuations 
is~$N, N-1, N-2, \dots$, while the spectrum for the deformed theory on the GGE side of the 
wall is~$N+1, N+2, \dots$, exactly as before.

However, not all these states are invariant under orientifolding. 
Recalling that the angular part of the wavefunction is of the form~$e^{i R \phi}$, 
and that the antipodal map \eqref{eq:involution} takes $\phi \mapsto \phi + \pi$, 
we see that the only wavefunctions which are invariant under the orientifold are the ones with $R\in 2\mathbb{Z}$. 
This means that for both $N = 2k+1$, and $N=2k$,  $k=1,2,\dots$,
the spectrum which survives the projection is 
$R = 2k, 2k-2,2k-4,\dots$ and 
$R = 2k+2,2k+4,\dots$ on the two sides, respectively, 
where $k = \lfloor\frac{N}{2}\rfloor$. 
Therefore, the index over the Hilbert space of two coincident maximal giants 
in AdS$_5\times \mathbb{RP}^5$ yields 
\begin{equation} \label{eqref:L-spectrum-two-giants-rp5}
    \mathrm{Tr}\,(-1)^Fq^R \=  \frac{q^{2k}}{1-q^{-2}} \= - \frac{q^{2k+2}}{1-q^2} \,.
\end{equation}
To summarize, one can think of the two coincident giants with their worldvolume identified 
as describing just one brane in the orientifold theory.
The spectrum of this brane then just consists of the states with wavefunctions that are invariant 
under the involution~\eqref{eq:involution}, 
which yields the two expansions in~\eqref{eqref:L-spectrum-two-giants-rp5} on the two sides of the wall.
It is the latter expansion that is relevant for the GGE.

\subsection{Topologically stable branes for~\texorpdfstring{$SO(2k)$}{SO(2k)}}
\label{sec:rigidbranes}

At~$\theta_0=\pi/2$, there is an invariant maximal giant with worldvolume 
topology~$S^3/\mathbb{Z}_2 = \mathbb{R} \mathbb{P}^3$ which forms the twisted sector 
of the $\frac{1}{2}$-BPS moduli space. 
As explained in \cite{Witten:1998xy}, a D3-brane can wrap the $ \mathbb{R} \mathbb{P}^3$ 
manifold only if the discrete torsions of the $B$ fields $B_{NSNS}$ and $B_{RR}$ both vanish. 
This means that such a configuration only exists for the theory dual to $
\mathcal{N}=4$ SYM with gauge group $SO(2k)$.

Thus, for the theory dual to~$\mathcal{N}=4$ SYM with gauge group $SO(2k)$, in addition 
to the previous dynamically stable supersymmetric brane 
configurations there is a topologically stable brane~\cite{Witten:1998xy,Aharony:2002nd}
given by a D3-brane wrapping the $3$-cycle  $\mathbb{RP}^3\subset \mathbb{RP}^5$. 
Since the homology group $H_3(\mathbb{RP}^5, \mathbb{Z}) = \mathbb{Z}_2$ has order two, 
a D3-brane wrapping the non-trivial cycle~$\mathbb{RP}^3$ 
represents a stable particle in AdS$_5$, whereas a configuration with two branes on the same 
cycle or a single brane wrapped twice is no longer topologically stable. 
The 5-form flux over $ \mathbb{RP}^5$ is just half that of the $S^5$ one \eqref{eq:5-form-flux-N}, i.e., 
\begin{equation} \label{eq:rp5-flux-k}
    \frac{1}{2\kappa^2_{10}T_3}\int_{\mathbb{RP}^5} F_{5} \= k \,. 
\end{equation}

These topologically stable branes only exist at the fixed point of the orientifold action~$x_1=x_2=0$. 
In particular, they do not admit any  fluctuations that deform the brane away from the fixed point and, 
in this sense, are rigid. This is to be contrasted with the supersymmetric maximal giants 
on a sphere, which admit supersymmetric fluctuations that take the giant away from the fixed point. 
In the Hamiltonian formalism, the wavefunction of any state (bosonic or fermionic) carrying non-zero 
angular momentum is peaked away from the origin, and therefore is not allowed.

Since the topologically stable brane is rigid, only its ground state contributes and there is no 
infinite series corresponding to it. The R-charge carried by the ground state is given by the 
flux over the D3-brane. The brane is mutually supersymmetric with respect to the pair of giants 
discussed above. Further, since its carries charge valued in the~$\mathbb{Z}_2$ homology,
it can either appear zero times or once. Therefore, the net contribution of the topologically
stable brane to the grand-canonical partition function is the multiplicative factor 
\be \label{eq:twistcontribution}
(1+q^k)\,,
\ee
which explains the corresponding factor in~\eqref{eq:gge-so2kintro}.

\bigskip

\section{Coincident maximal giants \label{sec:conincidentbranes}}

For $m$ coincident  maximal giants in AdS$_5\times S^5$, the bosonic sector is described by 
the matrix version of the Landau problem. The bosonic coordinates get promoted to 
Hermitian~$m\times m$ matrices that transform in the adjoint representation of the~$U(m)$ 
gauge symmetry group. Discarding the solenoid term as before, the Lagrangian is given by
\begin{equation} \label{eq:Lag-landau-matrices}
     \mathcal{L}^{(m)}_\text{Lan} \= \frac{N}{L} \, \mathrm{Tr}\Bigl( \, 
    \frac{1}{2} \, \bigl(\dot{X}^2+  \dot{Y}^2 \bigr) + \frac{1}{L}(X\dot Y - Y\dot X)   \Bigr) \,,
\end{equation}
where $X$ and $Y$ are hermitian $m\times m$ matrices.  

For the case of AdS$_5\times \mathbb{RP}^5$, in the untwisted sector we have~$m$ pairs 
of coincident maximal giants in AdS$_5\times S^5$ with the antipodal points of each pair of branes  
identified under orientifolding. 
As explained in the previous section, orientifolding implies that the orientation of the open strings 
stretching between the two antipodal points gets reversed. There are, as in the flat space orientfiold, 
two possible projections leading to the gauge groups~$SO$ and~$Sp$, respectively. 
We first focus on the  projection making the Hermitian matrices antisymmetric (and therefore purely 
imaginary), accordingly reducing the~$U(2m)$ symmetry group to~$SO(2m)$. 
We comment on the projection leading to the~$Sp(m)$ case at the end of this section. 

As for the case of one pair of giants, the Lagrangian without the solenoid term is 
\begin{equation} \label{eq:Lag-landau-m-pairs}
     \mathcal{L}^{m'}_\text{Lan} \= \frac{N'}{L} \, \mathrm{Tr} \Bigl( \,  
    \frac{1}{2} \, \bigl(\dot{X}^2+  \dot{Y}^2 \bigr) + \frac{B}{2}(X\dot Y - Y\dot X)   \Bigr) \,.
\end{equation}
Here we have used a notation in which we treat the~$U(N)$ theory and the~$SO(N)$ theory simultaneously.
For the~$U(N)$ theory, $N' = N$ is the 5-form flux through $S^5$, while for the the~$SO(N)$ 
theory~$N' = N/2$ is the 5-form flux through $\mathbb{RP}^5$. 
(The factor of~$1/2$ is due to the reduction of the volume by half.) 
The matrices~$X$ and~$Y$ are Hermitian matrices of size~$m'=m$ for the~$U(N)$ theory, 
while they are Hermitian and antisymmetric matrices of size~$m'=2m$
for the~$SO(N)$ theory.

The initial steps for the analysis of the matrix version of the Landau problem remain similar to the~$U(N)$ case. 
The Hamiltonian is  
\begin{equation}\label{Ham-2m}
     {H}^{m'}_\text{Lan}   \=  \mathrm{Tr}\left(\frac{L}{2N'} \bigl(P_X^2+P_Y^2 \bigr) \,+\, 
     \frac{N'}{2L}\frac{B^2}{4} \bigl(X^2 + Y^2 \bigr) 
     \,-\, \frac{B}{2} \bigl(X P_Y-Y P_X \bigr)\right) \,.
\end{equation}
After performing the change of coordinates
\begin{equation}
    Z \= \sqrt{\frac{N'}{2L}} \, \bigl(X+i Y \bigr)\, , \qquad 
    {Z}^\dagger \= \sqrt{\frac{N'}{2L}} \, \bigl(X-iY \bigr)\,,
\end{equation}
it takes the following form
\begin{equation}
    {H}^{m'}_\text{Lan} \= \sum_{i,j=1}^{m'} \Bigl( P_{ij}\overline{P}_{ij} + 
    \frac{B^2}{4}Z_{ij}\overline{Z}_{ij} - i\frac{B}{2}\left( Z_{ij}P_{ij} - \overline{Z}_{ij} \overline{P}_{ij} \right) \Bigr)\,,
\end{equation}
where $P_{ij} $ and $\overline{P}_{ij} $ are the conjugate momenta to $Z_{ij}$ and $\overline{Z}_{ij}$,
\begin{equation}
    P_{ij} \= -i \frac{\partial}{\partial Z_{ij}}\,,  \qquad 
    \overline{P}_{ij} \= -i \frac{\partial}{\partial \overline{Z}_{ij}}\,.
\end{equation}

The canonical commutation relations
\begin{equation}
    \bigl[ \, Z_{ij} \, , \, P_{ij} \, \bigr] \= i\,, \qquad
     \bigl[ \, \overline{Z}_{ij} \, , \, \overline{P}_{ij} \, \bigr] \= i\,, 
\end{equation}
imply that the operators
\begin{equation}
    {\Pi}_{ij} \= \overline{{P}}_{ij}-i\frac{B}{2}Z_{ij} \= -i \left( \frac{\partial}{\partial \overline{Z}_{ij}} 
    +\frac{B}{2}Z_{ij}\right) \= -i e^{-\frac{B}{2} Z_{ij}\overline{Z}_{ij}} \frac{\partial}{\partial  \overline{Z}_{ij}} 
    e^{\frac{B}{2} Z_{ij}\overline{Z}_{ij} } \,,
\end{equation}
\begin{equation}
    \overline{\Pi}_{ij} \= {{P}}_{ij}+i\frac{B}{2}\overline{Z}_{ij} \= -i \left( \frac{\partial}{\partial {Z}_{ij}} 
    -\frac{B}{2}\overline{Z}_{ij}\right) \= -i e^{\frac{B}{2} Z_{ij}\overline{Z}_{ij}} \frac{\partial}{\partial  {Z}_{ij}} 
    e^{-\frac{B}{2} Z_{ij}\overline{Z}_{ij}} 
\end{equation}
satisfy the following commutation relations, 
\begin{equation}
      \bigl[ \, {\Pi}_{ij} \,, \,  \overline{\Pi}_{ij} \bigr] \= B \,.
\end{equation}
The Hamiltonian then takes the form of the harmonic oscillator Hamiltonian in matrix form, 
\begin{equation} \label{hamiltonian-2m}
      {H}^{m'}_\text{Lan} \= \mathrm{Tr}\, \Pi^\dagger\,\Pi +  \frac12 {m'}^2B \,.
\end{equation}

\bigskip

The groundstate LLL wavefunctions are annihilated by the operators~$\Pi_{ij}$ or by the 
operators~$\overline{\Pi}_{ij}$ on the two sides of the wall, respectively. 
The corresponding bosonic groundstate wavefunctions take the form
\begin{equation} \label{eq:wavefunction-LLL-holo-antiholo}
     {\Psi}_{\text{LLL}}(Z,Z^\dagger) \= 
     \begin{cases}
     F(Z)\; \exp\bigl(- \frac{B}{2}\mathrm{Tr}\,\bigl(Z Z^\dagger\bigr) \bigr)
     \,, \quad \qquad & B > 0 \,, \\ 
     \overline{F}(\overline{Z})\; \exp\bigl(\frac{B}{2}\mathrm{Tr}\,\bigl(Z Z^\dagger\bigr) \bigr)\,,
     \quad \qquad & B < 0 \,, 
     \end{cases}
\end{equation}
where~$F$ and~$\overline{F}$ are holomorphic and antiholomorphic functions, respectively.  
The angular momentum operator in the symmetric gauge in the $Z,\overline{Z}$ coordinates is 
\begin{equation}
    \widehat{L} \= i\, C \sum_{i,j=1}^{m'} \left( Z_{ij}P_{ij} - \overline{Z}_{ij} \overline{P}_{ij} \right) 
    \=  C \sum_{i,j=1}^{m'}\left( Z_{ij}\frac{\partial}{\partial Z_{ij}} 
    - \overline{Z}_{ij}\frac{\partial}{\partial \overline{Z}_{ij}}\right) \,,
\end{equation} 
where~$C=1$ for the unitary group and~$C=1/2$ for the orthogonal group. 
This constant is introduced to account for the reduction of~$\widehat{L}$ by half  due to the 
orientifold as in \eqref{eq:L-angular-2}.

The Gaussian functions $\exp\bigl(\pm \frac{B}{2}\mathrm{Tr}\,\bigl(Z Z^\dagger\bigr) \bigr)$ 
are annihilated by~$\widehat{L}$,  
and therefore the angular momentum operator on the LLL wavefunctions gets reduced to its 
holomorphic or antiholomorphic parts acting on the functions $F(Z)$ or $\overline{F}(\overline{Z})$, 
respectively, as
\begin{equation}
    \widehat{L}  \, \bigr\vert_{\mathrm{LLL}} \=  
    \begin{cases}
         C \sum\limits_{i,j=1}^{m'}  Z_{ij}\dfrac{\partial}{\partial Z_{ij}} \,, & \quad B > 0 \,, \\ 
          \vspace{0.1cm} \\
         -C \sum\limits_{i,j=1}^{m'}  \overline{Z}_{ij}\dfrac{\partial}{\partial \overline{Z}_{ij}} \,, & \quad B < 0 \,.
    \end{cases}
\end{equation}

\bigskip

\subsection{Unitary gauge group}

Now it is convenient to separate the cases. Let us first consider the unitary gauge group. 
Depending on~$\text{sign}(B)$ the groundstate degeneracies depend on the matrix~$Z$ 
or~$\overline{Z}$. 
The wavefunctions are invariant under the group action and therefore the holomorphic 
(antiholomorphic) functions~$F(Z)$ ($\overline{F}(\overline{Z})$) are 
symmetric functions of the eigenvalues.\footnote{A complete set of invariants of a complex matrix~$Z$ 
under unitary equivalence is given by the traces of all possibles words made with products of~$Z$ 
and~$Z^\dagger$~\cite{PROCESI1976306}. The subset of holomorphic invariants is given by the 
traces of powers of~$Z$ or, equivalently, 
by symmetric functions of the eigenvalues.}  
Recall from our discussion of the supersymmetric Landau problem that the wavefunction is bosonic
 for~$B>0$ and fermionic for $B<0$. 
This means that the bosonic part of the wavefunction is symmetric for~$B>0$ and 
anti-symmetric for~$B<0$. In terms of the eigenvalues, the wavefunction for the groundstate 
takes the form
\begin{equation} \label{wf-LLL-2m}
    \Psi_{\text{LLL}}(Z,Z^\dagger) \= 
    \begin{cases}
        f(z) \,e^{-\frac{B}{2}\mathrm{Tr}\,\left(Z Z^\dagger\right)}\,, \quad \qquad & B >0 \,,\\
        \overline{f}(\overline{z}) \, \Delta(\overline{z})\,
        e^{\frac{B}{2}\mathrm{Tr}\,\left(Z Z^\dagger\right)} \,, \quad \qquad & B < 0 \,,
    \end{cases}
\end{equation}
where~$f$ and~$\overline{f}$ are symmetric functions of the eigenvalues, 
and~$\Delta(\overline{z}) = \prod_{1\leq i<j\leq m}(\overline{z}_i-\overline{z}_j)$ is the 
Vandermonde determinant.

In terms of the corresponding eigenvalues, we have 
\begin{equation} \label{eq:Leigenvalues}
    \widehat{L}  \, \bigr\vert_{\mathrm{LLL}} \=  
    \begin{cases}
         \sum\limits_{j=1}^m z_j\dfrac{\partial}{\partial z_{j}} \,, & \quad B > 0 \,, \\ 
         -\sum\limits_{j=1}^m \overline{z}_j\dfrac{\partial}{\partial \overline{z}_{j}} \,, & \quad B < 0 \,.
    \end{cases}
\end{equation}
In other words, on the space of symmetric polynomials, the eigenvalue~$\ell$ of~$\widehat{L}$ 
is simply the degree of the polynomial. 
The degeneracy of states as a function of the angular momentum is therefore given by the number of 
symmetric polynomials of~$z_i$ of degree equal to the angular momentum. 
One additive basis is the monomial symmetric polynomials in the $m$ variables~$z_i$ (monomials 
in any subset of the~$m$ variables, then symmetrized in all the~$m$ variables) 
of total degree~$\ell$, the number of which is given by the number of partitions of~$\ell$ into at 
most~$m$ parts. In order to relate the Landau problem spectrum to the giant graviton spectrum, 
we must label the states by their R-charge. 
A monomial in the variables $z_i$ of degree~$\ell$ has R-charge $-\ell$, and 
therefore the generating function counting the number of monomial symmetric polynomials in the 
variables $z_i$ as a function of the R-charge is 
\begin{equation}
     \prod_{r=1}^{m}  \frac{1}{1-q^{-r}} \,.
\end{equation}

\bigskip

For the other side of the wall, we have to include the R-charge on the Vandermonde $m(m-1)/2$ and the 
contribution of~$+1$ from each of the~$m$ fermionic degrees of freedom, thus totalling~$m(m+1)/2$ and 
the fermion number~$(-1)^m$. 
Finally, for one giant in AdS$_5\times S^5$, the inclusion of the effect of the solenoid term in the spectrum 
and the proper identification of the charges amounted to $R = N-\widehat{L}\vert_{\text{LLL}}$ 
where~$N$ is the flux through~$S^5$. 
For~$m$ giants, the R-charge is $mN-\widehat{L}\vert_{\text{LLL}}$.
Putting everything together, we obtain 
\begin{equation}
     \mathrm{Tr}\,(-1)^F \, q^R \, \Bigr|_\text{$m$ giants} \=
     \begin{cases}
     q^{mN} \displaystyle\prod\limits_{r=1}^{m}  \dfrac{1}{1-q^{-r}} \,,  \quad \qquad & B >0 \,, \qquad |q| > 1 \,,\\
     \vspace{0.1cm} \\
     (-1)^m q^{mN} q^{m(m+1)/2} \displaystyle\prod\limits_{r=1}^{m}  \dfrac{1}{1-q^{r}} \,, 
       \quad \qquad & B < 0 \,, \qquad |q| < 1 \,.
     \end{cases}
\end{equation}

\bigskip 

\subsection{Orthogonal projection}

Now we turn to the orthogonal gauge group. The difference in the treatment arises in the diagonalization 
of the operator $\widehat{L} $ over the groundstate wavefunctions. 
Now the matrices $Z, \overline{Z}$ are antisymmetric and the gauge symmetry is given by $SO(2m)$. 
A $2m\times 2m$ imaginary antisymmetric matrix $A$ can be reduced to the following form
\begin{equation}
    A \= \sigma_2 \otimes \mathrm{diag}(\lambda_1, \dots, \lambda_m) \,
\end{equation}
using $SO(2m)$ transformations $O$ which act as $A \to O A O^{-1}$, where 
the~$\pm \lambda_i \in \mathbb{R}$ are the real eigenvalues of $A$ and~$\sigma_2$ 
is the imaginary Pauli matrix. 
The eigenvalues of the antisymmetric complex matrices~$Z$ and~$\overline{Z}$ come 
in pairs of opposite sign which we denote by~$\pm z_i$ and~$\pm \overline{z}_i$, respectively.
The holomorphic invariants of a complex antisymmetric matrix~$Z$ of size~$2m$ under~$SO(2m)$ 
transformations are, as in Section~\ref{sec:boundary-indices}, given by 
$\mathrm{Tr}(Z^2), \mathrm{Tr}(Z^{4}),\dots$, and $ \mathrm{Pf}(Z)$.\footnote{Since $Z = X+i Y$ with 
$X,Y$ hermitian and antisymmetric (and therefore imaginary), any two holomorphic functions 
of~$Z$ which coincide on $X$ (or $iY$) are necessarily the same. Therefore, the function~$F(Z)$ 
is determined by its values for the real/imaginary part of~$Z$ and the holomorphic invariants for the 
complex matrix~$Z$ are the same as the ones computed in Section~\ref{sec:boundary-indices},
as in the unitary case.} In the eigenvalue basis these invariants are symmetric functions 
of~$z_i^2$,~$i=1,\dots,m$ and the Pfaffian~$\mathrm{Pf}(Z) = \prod\limits_{i=1}^m z_i$. 
Thus, the holomorphic function $F(Z)$ invariant under orthogonal transformations in the LLL 
wavefunctions \eqref{eq:wavefunction-LLL-holo-antiholo} is a symmetric function of the squared 
eigenvalues~$z_i^2$ and of~$\mathrm{Pf}(Z)$. With this notation, the operator~$\widehat{L}$ 
on the LLL is given by~\eqref{eq:Leigenvalues}.

The complete wavefunction is obtained by including the solenoid term. This solenoid flux 
through~$2m$ giants gets divided by two upon orientifolding and therefore the shift in 
the R-charge is~$2mN' = mN$ so that the total R-charge is~$mN-\widehat{L}\vert_{\text{LLL}}$. 
The total wavefunction takes this into account and can be effectively written as 
\begin{equation} \label{wf-LLL-2m-flux}
    \Psi_\text{(backgnd+LLL fluc)}(Z,Z^\dagger) \= 
        \left(\prod\limits_{i=1}^m z_i\right)^N  \; 
        \Psi_\text{LLL}(Z,Z^\dagger)
\end{equation}

\smallskip

Finally, we must only take into account the wavefunctions which are invariant under orientifolding. 
For the case of two coincident giants ($m=1$), the antipodal map acts as~$\varphi\mapsto \varphi+\pi$ 
in the coordinates~$(\rho,\varphi)$ or, equivalently, $z_1\mapsto -z_1$ in the complex coordinates. 
For general $m$, 
the wavefunction should be invariant under the transformation~$z_i \mapsto -z_i$, 
$i=1,\dots,m$. 
The Pfaffian is not invariant under orientifolding and therefore naively one would conclude that 
it plays no role in the counting. However, there is a subtlety in the orientifold theory depending 
on the parity of~$N$ because of the solenoid term, as follows. 

For $N$ even, the flux term is invariant under orientifolding and therefore the wavefunctions 
are purely symmetric functions of the squared eigenvalues (even powers of the Pfaffian are 
also invariant but they are also symmetric functions of the squared eigenvalues). 
In this case, the degeneracy of states associated to a given eigenvalue of~$R$ then arises completely 
from the symmetric function of the eigenvalues $z_i^2$,\footnote{One could also have obtained this 
statement directly by considering the groundstate wavefunctions for the unitary group and retaining 
the ones which are invariant under the projection~$z_i\mapsto -z_i$, $i=1,\dots,m$.} 
and it is given by the number of monomial symmetric polynomials of~$z^2_i$ of a given power 
corresponding to the eigenvalue. The groundstate degerenacies labelled by the R-charge are therefore given by
\begin{equation} \label{eq:degenarices-orthogonal-case-N-even}
\begin{split}
	  q^{mN}\prod_{r=1}^{m}\frac{1}{1-q^{-2r}}\,, \qquad & \text{$N$ even} \,,
\end{split}
\end{equation}
where the term $q^{mN}$ comes from the solenoid flux term in \eqref{wf-LLL-2m-flux}. 

For $N$ odd, the flux term under $z_i \mapsto -z_i$ yields a factor $(-1)^{N}$. Thus, for the wavefunctions 
to be invariant, the function $F(Z)$ must take the form a product of the Pfaffian $\mathrm{Pf}(Z)$ 
times symmetric functions of $z_i^2$. In this case, the previous counting yields,
\begin{equation} \label{eq:degenarices-orthogonal-case-N-odd}
\begin{split}
	  q^{mN}q^{-m}\prod_{r=1}^{m}\frac{1}{1-q^{-2r}}\,, \qquad & \text{$N$ odd} \,,
\end{split}
\end{equation}
where the factor $q^{-m}$ comes from the Pfaffian. We can express this answer uniformly 
for~$N=2k$ or~$N=2k+1$ as
\begin{equation} \label{eq:dynamical-d3-brane-contributions}
   \frac{1}{(q^{-2})_m}\,q^{2mk} \,.
\end{equation}
For the other side of the wall, we need to include the contribution of the Vandermonde 
determinant~$\Delta(\overline{z}^2)$ to take into account the antisymmetrization of the 
multi-eigenvalue bosonic determinant, as well as the fermions. The fermions effectively 
change~$N \to N+1$, interchanging the even and odd cases. Following the previous reasoning,
this leads to our final expression, 
for $N=2k$ or $N=2k+1$, 
\begin{equation} \label{eq:index-m-giants-orientifold}
     \mathrm{Tr}\,(-1)^F \, q^R \, \Bigr|_\text{$m$ giants} \=
     \begin{cases}
      \dfrac{1}{(q^{-2})_m} \, q^{2mk} \,, \quad \qquad & B >0 \,, \qquad |q|> 1 \,,\\
     \vspace{0.1cm} \\
     (-1)^m \, \dfrac{q^{2 \binom{m+1}{2}}}{(q^2)_m}\,q^{2mk} \,, \quad  \qquad & B < 0  \,,
     \qquad |q| < 1 \,.
     \end{cases}
\end{equation}

\subsection{Symplectic projection}

In the above derivation we assumed that the orientifold projection produces antisymmetric 
matrices, leading to a~$SO(2m)$ worldvolume gauge group. 
The other possibility, given by the choice~\eqref{eq:Msymp}, leads to the $Sp(m)$ gauge group, 
which arises when the boundary theory has a symplectic gauge group.

The derivation of the spectrum of fluctuations for the symplectic case follows the same steps as 
for the orthogonal one with minor modifications, as follows. 
The bosonic Lagrangian describing $2m$ coincident giants (ignoring the flux term) takes the same 
form as~\eqref{eq:Lag-landau-m-pairs}, 
where now the matrices $X,Y$ belong to the Lie algebra of $Sp(m)$, from which one can build 
the associated Hamiltonian. 
An element $A$ of the Lie algebra of $Sp(m)$ can be diagonalized using $Sp(m)$ transformations to take 
the form $A = \mathrm{diag}(\lambda_1, \dots, \lambda_m,-\lambda_1,\dots,-\lambda_m)$, 
where again the eigenvalues come in opposite pairs. 
On the physical side, the function $F(Z)$ must be a symmetric function of the squared 
eigenvalues~$z_i^2$ while on the GGE side the function~$\overline{F}(\overline{Z})$ is a symmetric 
function of $\overline{z}_i^2$ times the Vandermonde $\Delta(\overline{z}^2)$. Orientifolding again 
requires invariance under $z_i\mapsto -z_i$ (or $\overline{z}_i\mapsto -\overline{z}_i$) which does 
not affect the LLL wavefunction since there is no Pfaffian operator. 
We can write the wavefunction including the solenoid flux as~\eqref{wf-LLL-2m-flux}. 
The previous subtlety arising due to the parity of $N$, where $N$ is the number of D3-branes 
in the original $U(N)$ theory, does not appear in this case since the $Sp(k)$ theory can only arise 
as the orientifold of the $U(N)$ theory when $N=2k$.
Thus, the symplectic case is completely equivalent to the case of the orthogonal group with~$N$ 
even given by~\eqref{eq:degenarices-orthogonal-case-N-even}, and we 
recover~\eqref{eq:index-m-giants-orientifold} for the index. 
The $Sp(k)$ results are equal to the $SO(2k+1)$ results in the boundary theory, as consistent 
with the expectation from $S$-duality.

\subsection{The bulk giant graviton expansions}

The complete boundary index is recovered by summing over all contributions of $m$ pairs of 
maximal giants---with the addition of the twisted sector 
contributions when there is no discrete torsion---and multiplying the sum by 
the~$\frac{1}{2}$-BPS supergravity index in AdS$_5\times \mathbb{RP}^5$.
The latter consists of the index for states in AdS$_5\times S^5$ which are invariant under 
the orientifold projection. 

The spectrum of IIB supergravity in AdS$_5\times S^5$ compactified on $S^5$ was computed 
in~\cite{Kim:1985ez} and organized into representations of the superconformal algebra 
in~\cite{Gunaydin:1984fk}. The resulting \hbox{$\frac{1}{2}$-BPS} index for this spectrum, 
labelling states by one of the R-charges, takes the following form,
\begin{equation}
    I^{\mathrm{sugra}}_{\mathrm{ AdS}_5\times S^5}(q) \= \frac{1}{(q)_{\infty}}\,,
\end{equation}
which coincides with the~$N\to \infty$ index of~$\frac{1}{2}$-BPS states in~$U(N)$ $\mathcal{N}=4$ SYM. 
Indeed, the agreement for  large $N$  between the spectrum of Kaluza-Klein states and the spectrum 
of chiral operators of the conformal field theory was one of the first checks of the AdS/CFT 
correspondence~\cite{Witten:1998qj}.
To obtain the $\frac{1}{2}$-BPS index for type IIB supergravity in AdS$_5\times \mathbb{RP}^5$ 
it is enough to project to the invariant states under the identification~$x \sim -x$ for $x\in S^5$. 
The invariant single particle states under the identification are the ones which carry even values 
of the R-charge and therefore the complete~$\frac{1}{2}$-BPS index for type IIB supergravity 
in AdS$_5\times \mathbb{RP}^5$ takes the form
\begin{equation}
    I^{\mathrm{sugra}}_{\mathrm{ AdS}_5\times \mathbb{RP}^5}(q) \= \frac{1}{(q^2)_{\infty}}\,.
\end{equation}

Thus, the complete giant graviton expansion for the $\frac{1}{2}$-BPS index of type IIB string 
theory in AdS$_5\times \mathbb{RP}^5$ with~$k$ units of~$5$-form flux takes the following form
\begin{equation}
    I_{\mathrm{ AdS}_5\times \mathbb{RP}^5}(q) 
    \= \frac{1}{(q^2)_{\infty}} \sum_{m=0}^\infty (-1)^m \, \frac{q^{2 \binom{m+1}{2}}}{(q^2)_m} \, q^{2mk}\,,
\end{equation}
when there is no extra twisted sector, correctly reproducing the boundary indices for 
the~$SO(2k+1)$ and~$Sp(k)$ gauge theories~\eqref{eq:sp2k-so2k1-gge}.
When there is the extra twisted sector, we need to take into account the contribution
of the topologically stable branes~\eqref{eq:twistcontribution}. 
We obtain 
\begin{equation}
    I^{\,+ \,\mathrm{ twisted}}_{\mathrm{ AdS}_5\times \mathbb{RP}^5}(q) 
    \= \frac{1}{(q^2)_{\infty}} \sum_{m=0}^\infty (-1)^m \, \frac{q^{2 \binom{m+1}{2}}}{(q^2)_m} \, (1+q^k) \,q^{2mk}\,,
\end{equation}
which correctly reproduces the boundary index~\eqref{eq:gge-so2k} for the~$SO(2k)$ theory.

\medskip

Finally, we discuss the giant graviton expansion for the gauge group~$SU(N)$. 
As shown in Section \ref{sec:boundary-indices}, the giant graviton expansion for 
the~$\frac{1}{2}$-BPS index of $\mathcal{N}=4$ SYM with $SU(N)$  gauge 
group~\eqref{eq:GGE-SUN} is essentially the same as for the $U(N)$ gauge group, 
the only difference being in the supergravity ($I_\infty$) modes. The difference lies in the 
doubleton multiplet described in~\cite{Kim:1985ez}. This multiplet appears in the spectrum 
of Kaluza-Klein modes on~$S^5$ and can be gauged everywhere except at the boundary 
of AdS$_5$, thus defining a topological term. As explained in~\cite{Aharony:1999ti}, including 
it in the bulk corresponds to a dual $U(N)$ gauge group, while discarding it corresponds to 
the~$SU(N)$ case. Since the fields in the doubleton multiplet are those of the 
4-dimensional~$\mathcal{N}=4$ Maxwell multiplet, with index 
\begin{equation}
    I_{U(1)}(q) \= \frac{1}{1-q}\,,
\end{equation} 
this explains the difference between the two giant graviton expansions~\eqref{GGE_UNold} 
and \eqref{eq:GGE-SUN} for~$U(N)$ and~$SU(N)$, respectively.

\section*{Acknowledgements}

It is a pleasure to thank Ofer Aharony, 
Fri{\dh}rik Freyr Gautason, 
Yosuke Imamura, 
Ji-Hoon Lee, Douglas Stanford, Jesse van Muiden, 
and Edward Witten for useful and enjoyable discussions.
S.M.~acknowledges the support of the J.~Robert Oppenheimer 
Visiting Professorship at the Institute for Advanced Study, Princeton, USA and 
the STFC grants ST/T000759/1,  ST/X000753/1.
This work was supported by EPSRC grant no. EP/R014604/1.
G.E.~is supported by the STFC grant ST/V506771/1 and by an educational grant
offered by the A.~G.~Leventis Foundation.

\bigskip

\appendix

\section{Supersymmetry of giants in the Euclidean theory}

\label{sec:Appendix-killingspinors}

In this appendix we analyse the Killing spinor equations for AdS$_5\times S^5$ and show that in 
Euclidean signature with periodic time one can preserve half of the supersymmetries by 
performing a twist on the background. 
Equivalently, half of the spinors in the twisted Euclidean background are time independent and are 
well-defined in the geometry calculating the index.\footnote{A similar statement is true for supersymmetric 
AdS$_3 \times S^2$ as discussed in~\cite{Ciceri:2023mjl}.}
The introduction of a giant graviton in this background does not break further supersymmetries, 
since the Killing spinors which are independent of time are precisely the ones which survive 
the $\kappa$-symmetry projection (see Appendix A of the paper \cite{Gautason:2024nru}~v3 
for a similar derivation).

\subsection{Lorentzian \texorpdfstring{AdS$_5\times S^5$}{AdS5xS5} background}

This subsection is a brief review of~\cite{Grisaru:2000zn}. Consider the following spinor in AdS$_5\times S^5$,
\begin{equation} \label{eq:ads5s5-spinor}
\begin{split}
        \epsilon  & \=    e^{\frac{i}{2}\theta \gamma_5 \Gamma^\theta}e^{\frac{i}{2}\phi \gamma_5 \Gamma^\phi}
    e^{-\frac{1}{2}\chi_1  \Gamma^{\chi_1\theta}}
    e^{-\frac{1}{2}\chi_2  \Gamma^{\chi_2\chi_1}}
    e^{-\frac{1}{2}\chi_3  \Gamma^{\chi_3\chi_2}} \times 
    \\ & \qquad \quad  e^{\frac{i}{2}\mathrm{arcsinh}\left(\frac{r}{L}\right) \Gamma^r \gamma}
    e^{-\frac{i}{2L}t \Gamma^t\gamma }
    e^{-\frac{1}{2}\alpha_1 \Gamma^{\alpha_1 r}}
    e^{-\frac{1}{2}\alpha_2 \Gamma^{\alpha_2\alpha_1}}
    e^{-\frac{1}{2}\alpha_3 \Gamma^{\alpha_3\alpha_2}}
    \epsilon_0 \,,
\end{split}
\end{equation}
where the coordinates are the same as in the main text,~$\epsilon_0$ is an arbitrary constant spinor and 
the~$\Gamma$ matrices are orthonormal frame matrices belonging to the Clifford algebra~$\mathrm{Cl}(1,9)$ 
satisfying~$\{\Gamma_a,\Gamma_b\} = 2\eta_{ab}$. 
$\epsilon$ is a complex Weyl spinor which satisfies~$\Gamma^{11}\epsilon = \epsilon$, where~$\Gamma^{11} = \Gamma^{tr\alpha_1\alpha_2\alpha_3\theta\phi\chi_1\chi_2\chi_3}$. 
The other matrices appearing in the above expression  are~$\gamma= \Gamma^{t r\alpha_1\alpha_2\alpha_3}$ and~$\gamma_5 = \Gamma^{\theta\phi\chi_1\chi_2\chi_3}$. The spinor satisfies the Killing spinor equations given by the supersymmetry variation of the gravitino, 
\begin{equation} \label{eq:kse-ads5s5}
    D_M \epsilon - \frac{i}{1920} \Gamma^{PQRST}{\Gamma_M}F_{PQRST}\epsilon \= 0\,,
\end{equation}
where~$D_M$ is the covariant derivative on spinors,~$\Gamma^M$ satisfy~$\{\Gamma^M, \Gamma^N\} = 2 G^{MN}$, and the~$5$-form field strength is~$F = \frac{4}{L} \left( \varepsilon(\mathrm{AdS}_5) + \varepsilon(S^5) \right)$ with~$\varepsilon$ denoting the volume form. The equation \eqref{eq:kse-ads5s5} can be split into
\begin{equation} \label{eq:kse-ads5s5-split}
    D_\mu \epsilon - \frac{i}{2L}\gamma \Gamma_\mu \epsilon \=0\,, \qquad 
     D_m \epsilon - \frac{i}{2L}\gamma_5 \Gamma_m \epsilon \=0\,,
\end{equation}
where~$\mu$ runs over the AdS$_5$ coordinates and~$m$ over the~$S^5$ ones.
Since \eqref{eq:kse-ads5s5-split} is satisfied for any~$\epsilon_0$, the AdS$_5\times S^5$ background preserves 32 supersymmetries.

\subsection*{Giant gravitons in AdS$_5\times S^5$}

For the giant graviton D3-branes in this background, the residual supersymmetries are determined by imposing the following constraint on the Killing spinors on the brane worldvolume,
\begin{equation} \label{eq:kappa-sym-projection-epsilon}
    \Gamma \epsilon \= \epsilon\,,
\end{equation}
where 
\begin{equation}
    \Gamma = \frac{i}{4!} \varepsilon^{i_1\dots i_4}\partial_{i_1}X^{M_1}\dots \partial_{i_4}X^{M_4}\Gamma_{M_1\dots M_4}
    \end{equation}
is the matrix defining the $\kappa$-symmetry projection \cite{Grisaru:2000zn}. One can show that with the embedding \eqref{GGsol}, the condition \eqref{eq:kappa-sym-projection-epsilon} can be written as
\begin{equation} \label{eq:kappa-sym-cond}
    \bigl( \Gamma^{t\phi} +1 \bigr) \epsilon_0 \= 0\,.
\end{equation}
Since
\begin{equation}
    \mathrm{Tr}(\Gamma^{ t\phi }) \= 0 \, , \qquad (\Gamma^{ t\phi })^2 \= 1 \,,
\end{equation} 
the rotating giant gravitons preserve half of the background  supersymmetries.

\subsection{Twisted \texorpdfstring{AdS$_5\times S^5$}{AdS5xS5} background}

As explained in the text, we first focus on the  Euclidean bulk path integral corresponding to the index~$\text{Tr} \, (-1)^F \, e^{-\gamma (H-R)}$.
Accordingly, we perform a Wick rotation~$t=-i t_E$, 
make the identification~$t_E \sim t_E + \gamma$, and 
twist the~$\phi$-circle around the time coordinate. 
Since the radius of the Euclidean time circle is arbitrary, the only well-defined spinors in this geometry are independent of time.

First we impose a twist with an arbitrary parameter~$\omega$ through the following change of coordinates, 
\begin{equation}
    t' \= t \,, \qquad \phi' \= \phi - \omega t\,.
\end{equation}
The spinor which solves the Killing spinor equations now takes the form
\begin{equation} \label{eq:twisted-ads5s5-spinor}
\begin{split}
        \epsilon'  & \=   e^{\frac{i}{2}\theta \gamma_5 \Gamma^\theta}e^{\frac{i}{2}\phi' \gamma_5 \Gamma^{\phi}}e^{\frac{i}{2}\omega t' \gamma_5 \Gamma^{\phi}}
    e^{-\frac{1}{2}\chi_1  \Gamma^{\chi_1\theta}}
    e^{-\frac{1}{2}\chi_2  \Gamma^{\chi_2\chi_1}}
    e^{-\frac{1}{2}\chi_3  \Gamma^{\chi_3\chi_2}}\times
    \\ & \qquad \quad e^{\frac{i}{2}\mathrm{arcsinh}\left(\frac{r}{L}\right) \Gamma^r \gamma}
    e^{-\frac{i}{2L}t' \Gamma^{t}\gamma }
    e^{-\frac{1}{2}\alpha_1 \Gamma^{\alpha_1 r}}
    e^{-\frac{1}{2}\alpha_2 \Gamma^{\alpha_2\alpha_1}}
    e^{-\frac{1}{2}\alpha_3 \Gamma^{\alpha_3\alpha_2}}
    \epsilon_0' \,,
\end{split}
\end{equation}
where, again,~$\epsilon_0'$ is some arbitrary constant spinor and we have introduced 
the~$\Gamma$ matrices corresponding to the new frame. 
To simplify the exposition, from now on we omit the primed indices everywhere, so that every variable~$t,\phi$, their associated~$\Gamma$ matrices and the spinors in the rest of this subsection denote their primed versions. 
Demanding that the Killing spinor is time-independent yields
\begin{equation}
    \frac{\partial}{\partial t} \epsilon \= 0 \implies \Bigl(\omega \, \gamma_5 \, \Gamma^{\phi} - \frac{1}{L}\Gamma^{t}\gamma \Bigr) \epsilon_0 \= 0 \,,
\end{equation}
where we have commuted the matrices~$\gamma_5 \Gamma^\phi$ and~$\Gamma^{t}\gamma$ to the right. Using
\begin{equation}
    \Gamma^{11} \epsilon \= \epsilon \; \implies \; \Gamma^{11} \epsilon_0 \=  \epsilon_0 \,,
\end{equation}
we have
\begin{equation} \label{eq:projection-lorentzian-twisted-spinors}
    \Bigl(1  - \frac{1}{L\omega}\Gamma^{ \phi t} \Bigr) \epsilon_0 \= 0 \,.
\end{equation}
Since
\begin{equation}
    \mathrm{Tr}(\Gamma^{ \phi t}) \= 0 \, , \qquad (\Gamma^{ \phi t})^2 \= 1 \,,
\end{equation}
the equation~\eqref{eq:projection-lorentzian-twisted-spinors} only has solutions when $\omega =\pm \frac{1}{L}$, in which case it has 16 solutions. 

The value of~$\omega$ which produces the twist consistent with the index is~$\omega = \frac{1}{L}$. 
With this value, the rotating giant gravitons in AdS$_5\times S^5$ with embedding given in \eqref{GGsol} are static in the twisted background. 
One can check that the $\kappa$-symmetry projection \eqref{eq:kappa-sym-projection-epsilon} in the primed coordinates also leads to equation \eqref{eq:projection-lorentzian-twisted-spinors}. 
Thus, the Killing spinors which survive the projection due to the introduction of the giant gravitons 
in the background are precisely the time-independent Killing spinors.

\bibliographystyle{JHEP}

\bibliography{GGEbib}

\end{document}